\documentclass{article}
\usepackage{arxiv}

\usepackage[utf8]{inputenc} 
\usepackage[T1]{fontenc}    
\usepackage{hyperref}      
\usepackage{url}            
\usepackage{booktabs}       
\usepackage{amsfonts}       
\usepackage{nicefrac}       
\usepackage{microtype}      
\usepackage{lipsum}		
\usepackage{graphicx}
\usepackage{doi}

\title{GTOC 12: Results from TheAntipodes}

\date{}

\usepackage{authblk}

\author[1]{%
	{Roberto Armellin}%
}
\author[2]{%
	{Andrea Bellome}%
}
\author[3]{%
	{Xiaoyu Fu}%
}
\author[1]{%
	{Harry Holt}%
}
\author[1]{%
	{Cristina Parigini}%
}
\author[1]{%
	{Minduli Wijayatunga}%
}
\author[1]{%
	{Jack Yarndley\thanks{\texttt{jyar540@aucklanduni.ac.nz}}\hspace{1.2mm}}%
}

\affil[1]{Te P\=unaha \=Atea -- Space Institute, University of Auckland, Auckland 1010, New Zealand}
\affil[2]{ISAE-Supaero, Toulouse, 31400, France}
\affil[3]{University of Liverpool, Liverpool, L69 3BX, United Kingdom}

\renewcommand{\headeright}{}
\renewcommand{\undertitle}{Preprint}
\renewcommand{\shorttitle}{\textit{GTOC 12: Results from TheAntipodes}}

\hypersetup{
pdftitle={GTOC 12: Results from TheAntipodes},
pdfsubject={physics.space-ph},
pdfauthor={Roberto Armellin, Andrea Bellome, Xiaoyu Fu, Harry Holt, Cristina Parigini, Minduli Wijayatunga, Jack Yarndley},
pdfkeywords={trajectory optimization, space, asteroid mining, convex programming},
}

\usepackage{siunitx}
\usepackage{multirow} 
\usepackage[numbers]{natbib}
\usepackage{amsmath}
\usepackage{amssymb}
\usepackage{bm}
\usepackage{overcite}

\usepackage{subcaption}
\usepackage[version=4]{mhchem}
\usepackage{gensymb}
\usepackage{algorithm}
\usepackage[noend]{algpseudocode}
\usepackage{color,soul}
\usepackage[nocomma]{optidef}
\usepackage{tabularx,makecell}
\usepackage{tikz}
\usetikzlibrary{shapes.geometric, arrows, fit}
\usetikzlibrary{automata, positioning}
\usetikzlibrary{shapes.multipart}
\tikzset{
    startstop/.style =
        {rectangle, rounded corners, minimum width=3cm, minimum height=1cm,text centered, draw=black, fill=red!10, text width=3cm},
    io/.style = 
        {trapezium, trapezium left angle=70, trapezium right angle=110, minimum width=3cm, minimum height=1cm, text centered, text width=3cm,draw=black, fill=blue!30},
    process/.style = {rectangle, rounded corners, minimum width=3cm, minimum height=1cm, text centered, text width=3cm, draw=black, fill=orange!10},
    decision/.style = {diamond, minimum width=3cm, minimum height=1cm, text centered, draw=black, fill=green!10, text width=3cm, inner sep=-8pt, aspect=1.3},
    arrow/.style = {thick,->,>=stealth},
    myfit/.style={draw,dashed,black, inner xsep=15pt, inner ysep=20pt, rounded corners=5pt},
    mytitle/.style={draw,densely dashed,black, fill=black!10, inner sep=5pt, right, xshift=10pt}
}

\usepackage{adjustbox}

\begin{document}
\maketitle

\begin{abstract}
    We present the solution approach developed by the team `TheAntipodes' during the 12th edition of the Global Trajectory Optimization Competition (GTOC 12). An overview of the approach is as follows: (1) generate asteroid subsets, (2) chain building with beam search, (3) convex low-thrust trajectory optimization, (4) manual refinement of rendezvous times, and (5) optimal solution set selection. The generation of asteroid subsets involves a heuristic process to find sets of asteroids that are likely to permit high-scoring asteroid chains. Asteroid sequences `chains' are built within each subset through a beam search based on Lambert transfers. Low-thrust trajectory optimization involves the use of sequential convex programming (SCP), where a specialized formulation finds the mass-optimal control for each ship's trajectory within seconds. Once a feasible trajectory has been found, the rendezvous times are manually refined with the aid of the control profile from the optimal solution. Each ship's individual solution is then placed into a pool where the feasible set that maximizes the final score is extracted using a genetic algorithm. Our final submitted solution placed fifth with a score of $15,489$.
\end{abstract}


\section{\label{sec:Part1}Introduction}

Global Trajectory Optimization Competition (GTOC) is an international competition first proposed by the Advanced Concepts Team (ACT) at ESA in 2005 to challenge the international scientific community with ‘nearly impossible problems’ in the field of space trajectory optimization \citep{Izzo2007}. The competition entails planning multi-target missions, where a single spacecraft or a fleet of spacecraft are launched, each visiting multiple orbital way-points. 

In the 12th edition (GTOC 12), several mining ships that are launched from Earth are expected to rendezvous with multiple asteroids to mine them and return the highest possible mined mass to Earth \citep{gtoc12ProblemDescription}. The design of this mission is complicated by the fact that the visiting order of targets is not known a priori but is an integral part of the optimization itself. This results in complex mixed-integer non-linear programming (MINLP) problems \citep{Schlueter2013}, which are also known in the literature as hybrid-optimal control problems (HOCP) \citep{Ross2005}. In MINLP/HOCP, the combinatorial problem aims to find the optimal sequence of targets to be visited. It is mixed with optimal control theory to identify one or more locally optimal trajectories for a candidate sequence in terms of launch date, phasing between targets, and thrust arcs. In GTOC, the added complexity of designing multi-target missions mainly arises from two principles \citep{Izzo2007, Torn1999}. Firstly, the mixed-integer domain is large, such that multiple local optima exist. Secondly, objective functions and/or constraints are unconventional, such that no standard methods or existing software can solve the problem by operating as a black box. For instance, in GTOC 12, the released dataset contains 60,000 asteroids, and the mission's overall duration extends up to 15 years. The objective function is influenced by the number of mining ships, a value determined through the optimization process itself \citep{gtoc12ProblemDescription}.

Hence, it is impractical to refine and optimize all possible trajectories that might contribute to the solution, and the key to solving this class of problems lies in managing this complexity efficiently. A common way employed by many teams in past editions to handle this complexity is represented by multi-fidelity approaches \citep{DiCarlo2018, bellome2023phd, BELLOME2023}, where approximate transfer analyses estimate crucial parameters for any target-to-target transfer, such as the propellant consumption, and/or the time of flight. In this way, one first uses low-fidelity models to quickly build multi-target trajectories (solving the combinatorial problem), and then more detailed models are employed to reconstruct the actual trajectory using the low-fidelity solution as an initial guess (solving the optimal control problem). For example, Edelbaum-like approaches \citep{edelbaum1961propulsion} are typically used in GTOC to approximate low-thrust transfers \citep{act2022gtoc, armellin2022team}. Approximate methods might also be used to estimate $ \Delta v$ and time of flight in complex dynamical environments as in GTOC 9 \citep{shen2021simple}. Machine learning approaches have also been employed to quickly provide approximate transfer information and ease the transition between the combinatorial problem and the refinement \citep{li2020deep, act2022gtoc}.

\begin{figure}[hbt!]
    \centering
    \begin{tikzpicture}[node distance=2.0cm]
     
    \node (subset_generation) [process] {Subset generation};
    \node (beam_search) [process, below of=subset_generation] {Beam search};
    \node (beam_search_outcome) [startstop, below of=beam_search] {Self-cleaning \\ sequences};
       \node [inner xsep=1em, inner ysep=1em, fit=(beam_search_outcome) (subset_generation)(beam_search),draw,dotted,black] (box) {};
       
    \node (low_thrust) [process, right of=subset_generation,xshift=3cm] {Sequential Convex \\ Optimization };
    \node (manual_ref) [process, below of=low_thrust] {Manual refinement};
    \node (solution_pool) [startstop, below of=manual_ref] {Solution pool};
    \node (mip_selection) [process, right of=manual_ref, xshift=3cm] {Optimal solution \\ set selection};
    \node (final_solution) [startstop, below of=mip_selection] {Combined solution};

    \node [inner xsep=1em, inner ysep=1em, fit=(low_thrust) (manual_ref) (solution_pool),draw,dotted,black] (box1) {};

    \node[fill=white] at (box.north) {Mission Design};
    \node[fill=white] at (box1.north) {Low-Thrust Optimization};

    \draw [arrow] (box) -- (box1);
      \draw [arrow] (box1) -- (mip_selection);
    
    \draw [arrow] (subset_generation) -- (beam_search);
    \draw [arrow] (beam_search) -- (beam_search_outcome);
    \draw [arrow] ([xshift=2mm] low_thrust.south) -- ([xshift=2mm] manual_ref.north);
    \draw [arrow] ([xshift=-2mm] manual_ref.north) -- ([xshift=-2mm] low_thrust.south);

    \draw [arrow] (manual_ref) -- (solution_pool);
    \draw [arrow] (mip_selection) -- (final_solution);
    
    \end{tikzpicture}
    \caption{Schematic of the solution pipeline developed by team `TheAntipodes' for GTOC 12.}
    \label{fig:mainschematic}
\end{figure}
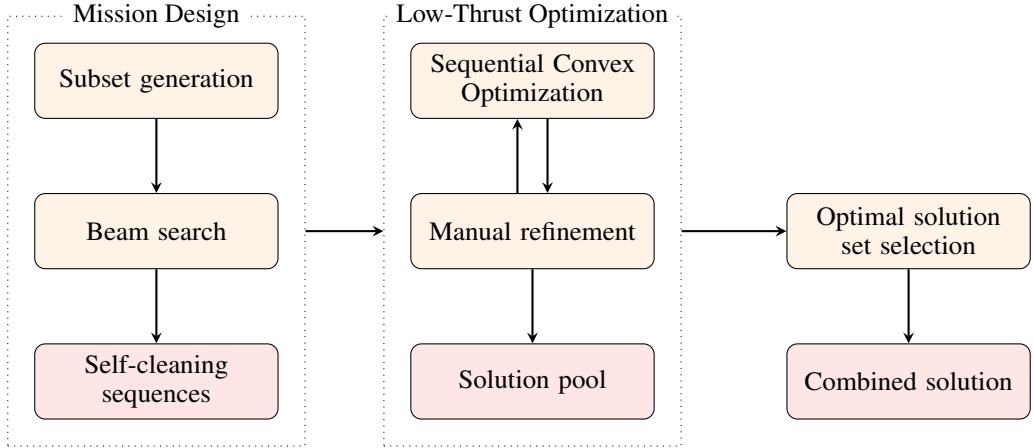

A multi-fidelity approach has been implemented by the team `TheAntipodes' to tackle the problem presented in GTOC 12. The primary steps of this approach are presented in Figure~\ref{fig:mainschematic}. The approach is based on five main computational blocks (with a light yellow background in the schematic): (1) asteroid subset generation, (2) chain building via beam search, (3) convex low-thrust trajectory optimization, (4) manual rendezvous time refinement, and (5) optimal solution set selection. Blocks (1) and (2) are considered as the mission design phase of the solution generation, whilst blocks (3) and (4) are conducted iteratively to perform the low-thrust trajectory optimization. Output blocks, represented by light red blocks, naturally provide this separation of each phase of the solution approach. These are: (1) sequences of asteroids permitting high-quality `self-cleaning' sequences, (2) a pool of refined low-thrust ship trajectories, and (3) the combined solution for submission (the output of the whole approach). 

Using this methodology, the team obtained the fifth-best solution of the competition with a score of $J = 15,488.896$. All key component blocks were developed from scratch during the competition, except basic astrodynamics libraries such as ephemerides, orbital element conversions, and a Lambert solver. Third-party solvers and optimization routines were then utilized to solve the formulated problems.

The paper is structured as follows. Firstly, our preliminary investigations of the GTOC 12 problem are presented in Section ~\ref{sec:ProblemAnalysis}. Section~\ref{sec:Subsets} covers the pruning methods used on the asteroid dataset to retain only the asteroids that can permit efficient transfers to/from Earth and the subsequent division of these into independent pools (`subsets'). These asteroid pools are provided as input to a beam search algorithm in Section~\ref{sec:BeamSearchAndrea}, to generate efficient asteroid sequences (`chains') that deploy miners and collect material from the same set of asteroids (referred to as self-cleaning). The mass-optimal control profile for these asteroid sequences is found using sequential convex programming (SCP), discussed in Section~\ref{sec:SCP}. A manual refinement of the rendezvous times is then performed (Section~\ref{sec:ManualRefinement}) to improve their returned material, which, in combination with the beam search, leads to the creation of a pool of solutions in Section~\ref{sec:PoolSolutions}. Section~\ref{sec:Optimiser} details the selection of the set of ships from the pool, which maximizes returned material and takes the bonus factor into account. Section~\ref{sec:Performance} provides a brief overview of the computational performance of the solution approach, and Section~\ref{sec:FResult} provides details of our optimal solution, including trajectory visualizations. Finally, our conclusions and limitations of the approach are presented in Section~\ref{sec:Conclusion}.

\section{\label{sec:ProblemAnalysis}Problem Analysis and Initial Investigations}

As mentioned in \citep{gtoc12ProblemDescription}, the objective of GTOC 12 is to maximize the total resource value $J$ mined from the asteroids. This value is determined by the total mined mass, which is weighted by an asteroid-dependent bonus coefficient $\beta$. Since the mined mass is proportional to the time spent on the asteroids, the optimal strategy involves missions that depart from Earth as soon as possible and return as late as possible. Additionally, it is crucial to minimize transfer times between asteroids to deploy miners as soon as possible and collect mined mass as late as possible. These considerations played a pivotal role in both the subset generation and the configuration of beam search parameters.

\subsection{\label{sec:ObjectiveFunction}Objective Function Analysis}
\begin{figure}[hbt!]
    \centering
    \includegraphics[width = 16cm]{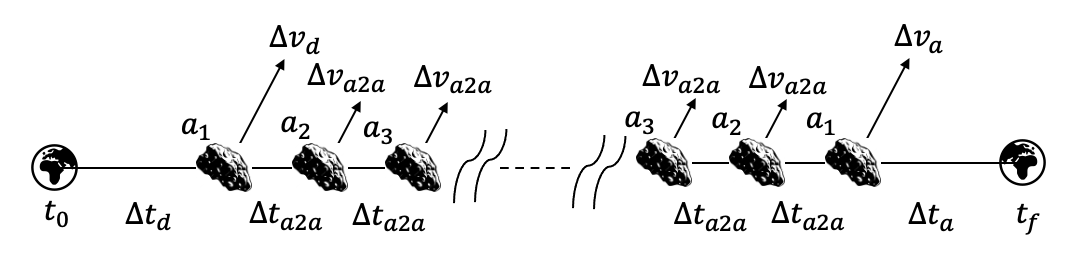}
    \caption{Schematic of the simplified mission.}
    \label{fig:simplifiedSol}
\end{figure}

Based on the considerations above, a simple analysis was carried out to understand the problem. The analysis was based on the following simplifying assumptions: 
\begin{enumerate}
\item The solution consists only of identical missions.
\item A ship deploys $I$ miners, collects from $I$ asteroids, and returns to Earth. 
\item The mission timeline is as provided in Figure~\ref{fig:simplifiedSol}, where $t_0$ is the departure epoch from Earth (64,328 MJD), and $t_f$ is the epoch of return to the Earth (69,807 MJD), for a mission duration of 15 years. $\Delta t_{d/a}$ is the time of flight of the departure/arrival leg from/to the Earth with $\Delta t_{a} = \Delta t_{d}$, $\Delta t_{a2a}$ is the time of flight of the transfer between asteroids, assumed to be the same for all transfers. 
\item Each transfer cost is represented by a $\Delta v$, as shown in Figure~\ref{fig:simplifiedSol}. The cost of the first/last arc is labeled with $\Delta v_{d/a}$. It is assumed $\Delta v_{d} = \Delta v_{a}$. All the asteroid-to-asteroid legs have a cost $\Delta v_{a2a}$. However, the intermediate transfer from the final departure leg to the first arrival leg has $\Delta v_{a2a} = 0$, as we assumed collection in reverse order to limit the propellant consumption (while the collection order has no impact in terms of total collected mass, collecting in reverse order entails a minimization of the propellant, as the largest mined mass is carried onboard for a minimal time). 
\end{enumerate}
Using these assumptions, the computation of the solution score $J$ (assuming no bonus penalization) requires only four parameters, $\Delta t_{d/a}$, $\Delta t_{a2a}$, $\Delta v_{d/a}$ and $\Delta v_{a2a}$. The procedure to evaluate a solution is provided in Algorithm~\ref{algo:simple}.

\begin{algorithm}[hbt!]
  \caption{Simplified GTOC 12 solution score calculation}\label{algo:simple}
  \begin{algorithmic}[1]

     \State Set estimated parameter values, $\Delta t_{d/a}, \Delta t_{a2a}, \Delta v_{d/a}, \Delta v_{a2a}$
     \State Set the asteroid mining rate, $k = 10$ kg/yr, and dry mass $m_d = 500$ kg
     \State Compute the maximum number of miners, $I_{\max} =  \textrm{floor} \left(\frac{(t_f-t_0)/2 - \Delta t_{d} - \Delta t_{a}}{\Delta t_{a2a}}\right)+1$
    \State Compute the fixed time grid for transfers of length $2I_{\text{max}} + 2$, \\ \quad ${\bf t} = [t_0,t_0+\Delta t_{d} + 0\Delta t_{a2a}, \dots, t_0+\Delta t_{d}+(I_{\max} - 1)\Delta t_{a2a},$\\ \quad \quad \quad \quad $ t_f-\Delta t_{a}-(I_{\max} - 1)\Delta t_{a2a},\dots, t_f-\Delta t_{a}-0\Delta t_{a2a}, t_f]$
     
     \For {$i=1,\dots,I_{\max}$} 
    \State Set mined mass for using $i$ miners, $M_{i} = 0$
     \For{$j=1,\dots,i$} 
     \State Update mined mass according to the fixed time grid, $M_i = M_i + k [\mathbf{t}_{2I_{\text{max}}+2-j} - \mathbf{t}_{j+1}]$ 
     \EndFor 
     \State Set spacecraft final mass $m_{i,f} = m_d + M_{i}$ 
     \State Compute the spacecraft initial mass $m_{i,0}$ backward starting from $m_{i,f}$ adding the propellant mass via the rocket equations, removing mass at collections, and adding mass at deployments
     \If {$m_{i,0} \ge 3,000$ kg }
     \State Find reduction factor $\alpha <1 $ for $M_i$ such that $m_{i,0} = 3,000$ kg
     \State Update mined mass, $M_{i} = \alpha M_{i}$
     \EndIf
     \EndFor
     \State Set $j$ to the $i$ that maximizes $M_i$
     \State Compute permitted number of ships, $N = \textrm{floor}(\min(100,2\exp(\rho M_j)))$
     \State Compute the total mined mass (objective value), $J = N M_j$  
     \State Output $j, N, J, m_{j, 0} $  
   \end{algorithmic} 
\end{algorithm}

\begin{figure}[]
    \centering
    \includegraphics[width = 16cm]{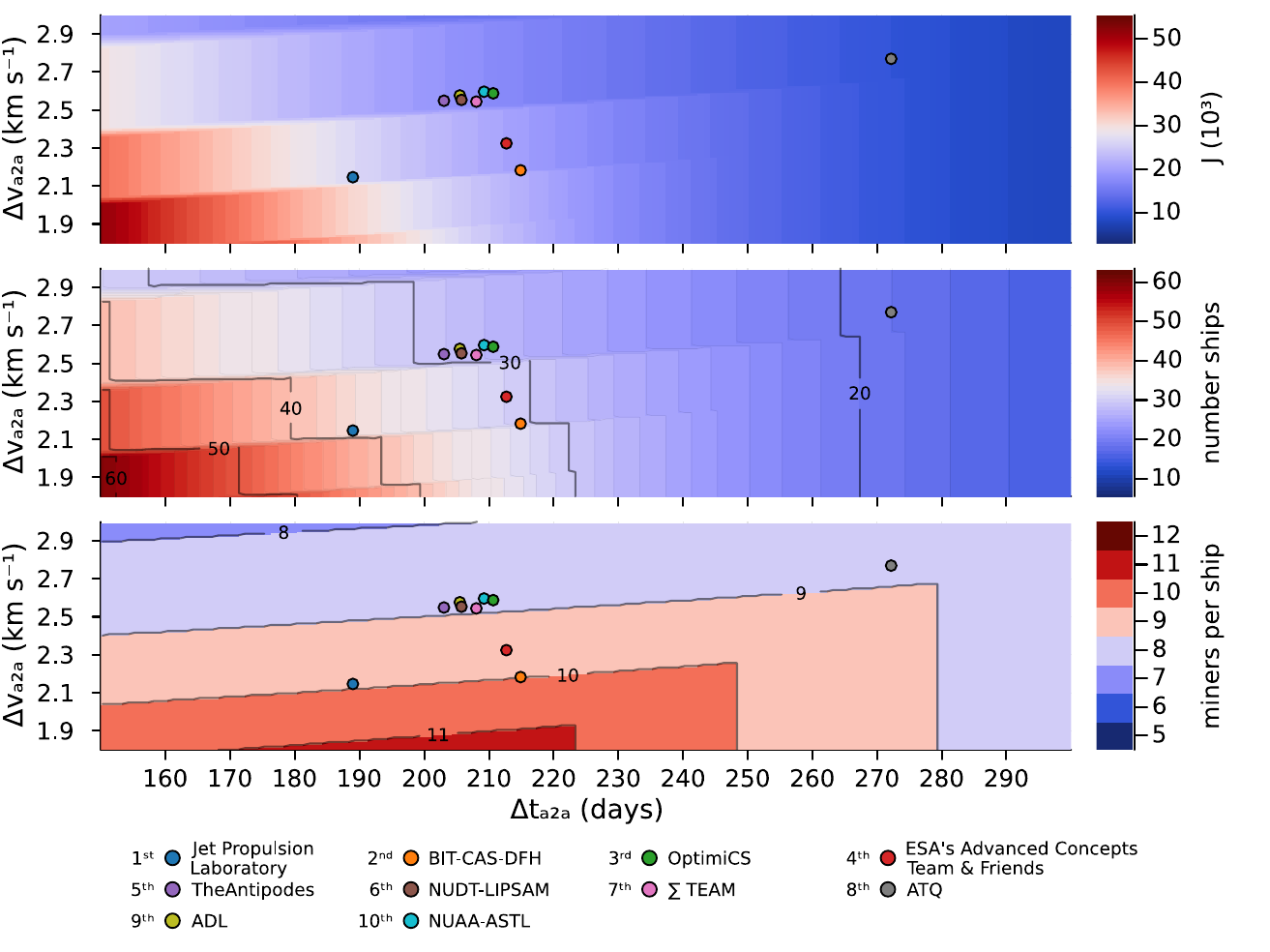}
    \caption{Preliminary analysis of the objective function for $\Delta t_{d/a}= 500$ days and $\Delta v_{d/a} = \SI{6}{\kilo\meter\per\second}$. Markers show how this compares to the final submitted solutions of the top 10 teams in the GTOC 12 competition.}
  \label{fig:prelAnalysis}
\end{figure}

To start building some understanding of the problem, we decided to assume reasonable values for $\Delta t_{d/a}, \Delta v_{d/a}$ and performed a parametric analysis on $\Delta t_{a2a}, \Delta v_{a2a}$. In particular, we assumed $\Delta t_{d/a}= 500$ days and $\Delta v_{d/a} = \SI{6}{\kilo\meter\per\second}$. The parametric analysis was performed with $\Delta t_{a2a} \in [150, 300]$ days and $\Delta v_{a2a} \in [1.8, 3.0] ~\SI{}{\kilo\meter\per\second}$. The values of $\Delta t_{d/a}$ and $\Delta v_{d/a}$ were taken from a preliminary assessment of low-thrust trajectories in the main asteroid belt, whereas the ranges for $\Delta t_{a2a}$ and $\Delta v_{a2a}$ were obtained from investigations on Lambert transfers between asteroids. Figure~\ref{fig:prelAnalysis} summarizes the main results of this analysis, providing valuable insights into the problem. 

It is evident that for longer transfer times, the cost of transfers is less important as the number of miners and the mined mass decrease sufficiently to make the mission feasible. Conversely, for shorter transfer times, transfer costs become more critical because only efficient transfers allow the deployment of more miners and the collection of more mined mass. Shorter transfer times are crucial for maximizing the score due to the direct impact on the number of miners and collected mass. 

Importantly, changes in the optimal number of miners directly impact the final score and the allowed number of ships (as reflected in the shape of the contour levels in all three plots). Interestingly, for a fixed $\Delta v_{a2a}$, it is not always the case that the optimal number of miners increases with the decrease in time of flight. A shorter time-of-flight causes more mined mass to be collected, so the number of miners may need to be reduced in order to support the larger mass pickup. Overall, this analysis supports the observation that mining around ten asteroids per ship is sufficient to achieve high scores. It is clear that whilst reaching the limit of 100 ships appears unfeasible, solutions between 20 and 40 ships are clearly feasible. Thus, a total $\Delta v$ of around $\SI{45}{\kilo\meter\per\second}$ per ship appeared to be a reasonable value for the beam search analysis. 

Lastly, this analysis is presented in combination with the final values derived from the solutions submitted by the top 10 teams in GTOC 12. Comparing the estimates with actual values from the teams' solutions shows this preliminary investigation to be quite accurate (with some exceptions like OptimCS and ATQ).

\subsection{Main Design Decisions} 

GTOC 12 introduced some interesting features that made the problem extremely challenging:
\begin{enumerate}
    \item The number of ships is not known a priori, but is a result of the performance of the global solution.
    \item Missions can be coupled (`mixed') because the deployment of a miner and the collection of the mined mass for a single asteroid can be done by different missions.
    \item The total resource value is influenced by other teams' submissions via the asteroid-specific bonus.
\end{enumerate}
To achieve a solution within the limited time frame available, the team decided to prioritize strategies that would produce sub-optimal solutions but with the advantage of a drastic reduction in problem complexity. 

Leveraging the preliminary analysis of the objective function, it was decided to focus primarily on self-cleaning missions. This decision was based on the fact that it is possible to find groups with eight or more asteroids with similar orbital parameters (permitting low-cost transfers throughout the problem timeframe), enabling the creation of good-scoring self-cleaning missions. The option for a mixed solution remained as a possible improvement of the score, but around two weeks into the competition, it was decided to avoid the difficulties of a fully mixed design. This means that all missions could be optimized separately, and their combination only performed as a final process before solution submission.

A second decision made was to neglect the impact of the asteroid bonus during the optimization of the missions. The impacts of the bonus would have a much higher impact on the selection of missions compared to the decisions made within a mission (such as how long to stay at an asteroid). Thus, the bonus was considered only in the final selection process. Thirdly, after a large amount of analysis, gravity assists were considered detrimental due to their general increase in time-of-flight outweighing the relatively minor decrease in transfer cost.

\section{\label{sec:MissionDesign}Mission Design}
The process to establish the asteroid rendezvous sequences and computing approximate rendezvous times is described. This process is split into two parts: the generation of the subsets to reduce the problem size and the implementation of two beam search variations to select efficient asteroid sequences from these subsets.

\subsection{\label{sec:Subsets}Subset Generation}
The size of the asteroid set (there are $60,000$ asteroids in the dataset) makes it challenging to get suitable results from the beam search, especially with the goal of finding self-cleaning solutions. A solution to this was proposed by the creation of `subsets': small groups of asteroids that permit low Lambert $\Delta v$ transfers between all members along the time frame of the GTOC 12 problem. The creation of these subsets enabled the development of high-quality self-cleaning asteroid chains from an adapted beam search algorithm. 

Firstly, the asteroid set was pruned to remove those that do not have characteristics that permit efficient transfer from Earth. A loose pruning cut to the asteroid dataset was applied as follows:
\begin{align}
    a \geq 4.0 \lor e \geq 0.4 \lor i \geq 15
\end{align}
where $a$ is the semi-major axis of the orbit non-dimensionalized to 1 AU, $e$ the eccentricity and $i$ the inclination of the orbit in degrees. This pruning step removed approximately $20\%$ of the asteroids. Next, the pruned set of asteroids is grouped according to a statistic we refer to as the \textit{average} Lambert $\Delta v$. 
\begin{figure}[hbt!]
    \centering
    \includegraphics[width = 16cm]{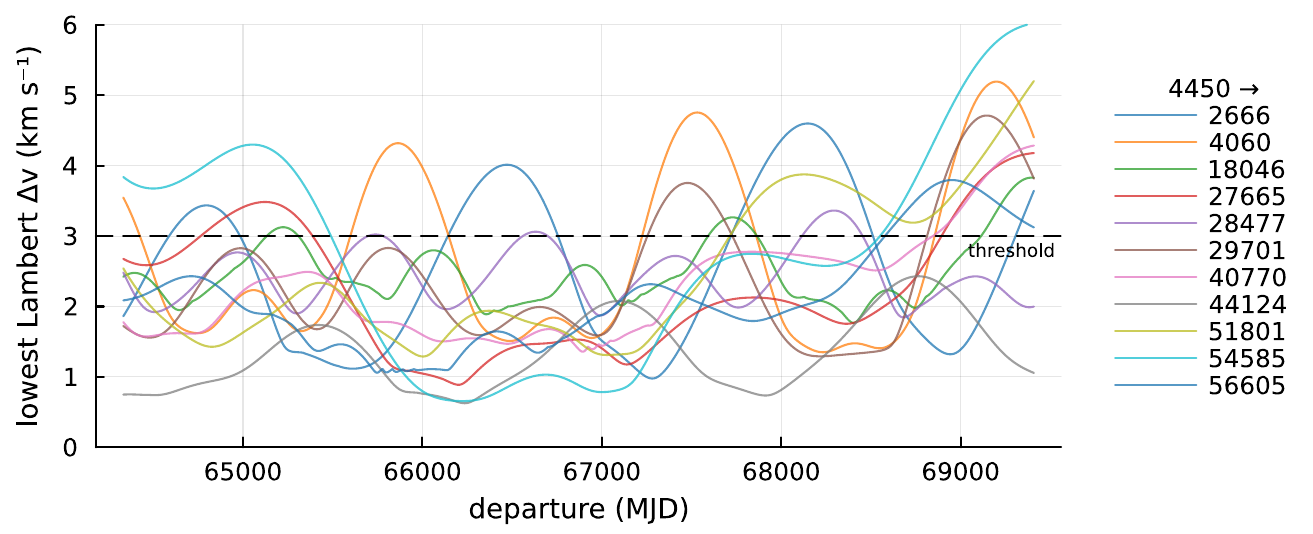}
    \caption{Lowest Lambert $\Delta v$ from asteroid $4450$ to other asteroids which meet the average threshold of $< \SI{3}{\kilo\meter\per\second}$.}
    \label{fig:lambert_average}
\end{figure}

The \textit{average} Lambert $\Delta v$ is calculated from every asteroid in the pruned set to all others by averaging an estimate of the lowest Lambert $\Delta v$ every $5$ days throughout the permitted duration of GTOC 12 (note \textit{average} is referring to averaging across time, not destination). To estimate the lowest Lambert $\Delta v$, the minimum $\Delta v$ Lambert transfer to the target is taken across a range of time-of-flights ranging from 150 to 400 days with 25-day increments. A plot of how this changes over time for an individual asteroid solution can be seen in Figure~\ref{fig:lambert_average}. In order to improve computational performance, a pruning step is added which prevents the calculation of this statistic for transfers that are highly likely to be expensive. Transfers have a Lambert $\Delta v$ greater than $\SI{10}{\kilo\meter\per\second}$ from $65,000$ MJD to $65,200$ MJD are not considered.

A subset is then generated with all asteroids that meet the threshold of \textit{average} Lambert $\Delta v \leq \SI{3}{\kilo\meter\per\second}$ in combination with the departure asteroid. The $\SI{3}{\kilo\meter\per\second}$ parameter was set based on manual tuning and acts as a trade-off parameter between group size and efficiency in accordance with the preliminary analysis. This process is repeated starting from every asteroid within the pruned group. Thus, every asteroid in the pruned set will have a specific set that meets this criterion.

\begin{figure}[hbt!]
    \centering
    \includegraphics[width = 16cm]{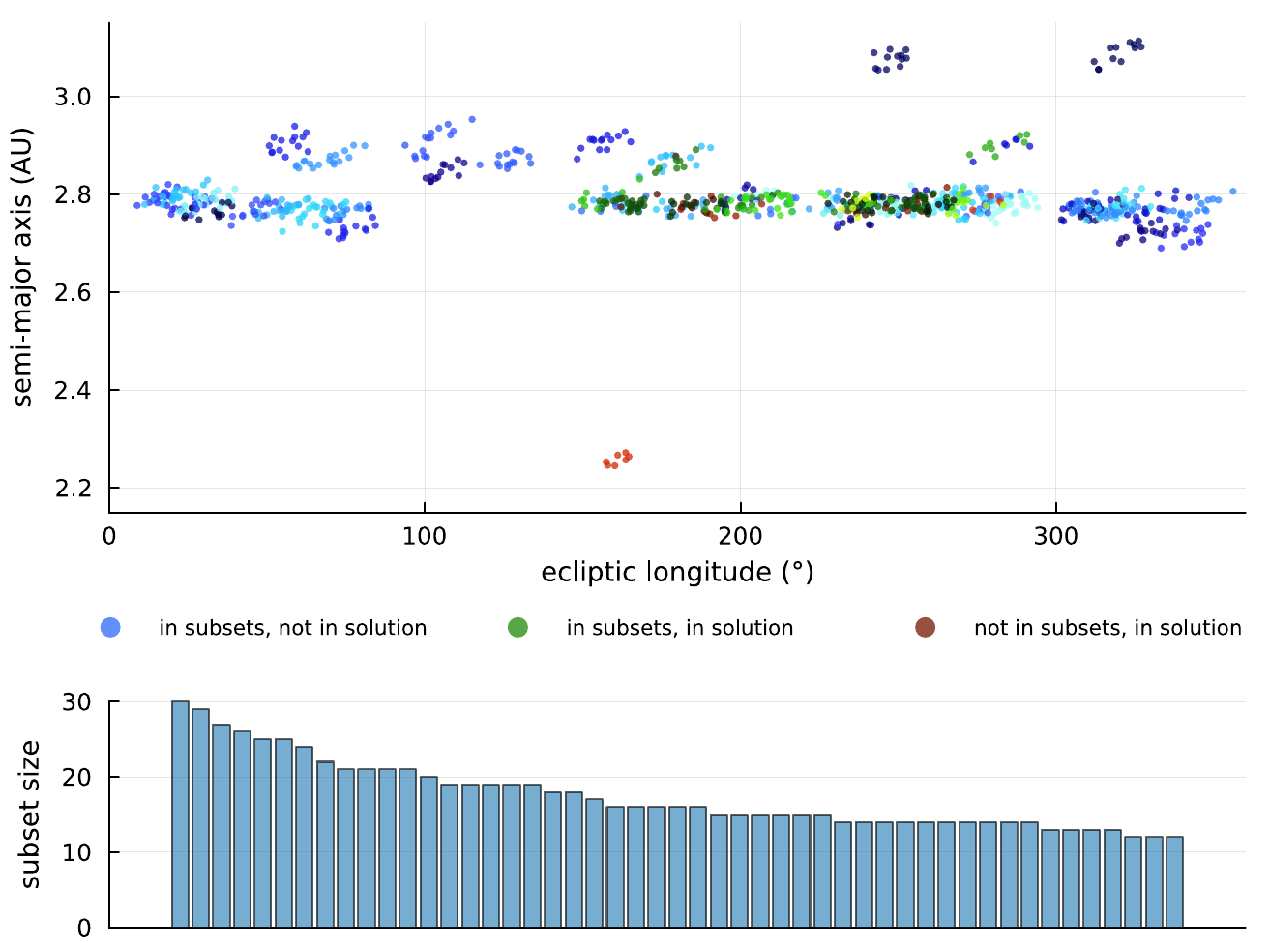}
    \caption{Details on the use of independent subsets within the submitted solution. (Top) A comparison of the semi-major axis and ecliptic longitude for the identified subsets and the submitted solution, with shades of color representing the different sets of asteroids for each category. (Bottom) The sizes of the independent subsets used to generate the majority of the solution. The reference time is $65,000$ MJD, approximately when the first asteroid rendezvous occurs.}
    \label{fig:subset_classical}
\end{figure}

Each subset identified with the above procedure has no guarantee of independence, meaning the same asteroid can be part of multiple subsets. Therefore, we studied several methods to merge these subsets to identify larger but independent subsets of asteroids, referred to as independent subsets. In this way, solutions with asteroids of different independent subsets are automatically compatible. The first method implemented was a greedy approach: ordering all of the sets by size and iteratively removing any duplicated asteroids in the largest set from all smaller subsets. In this process, a reasonable number of independent subsets (30-50) with more than ten members were created. These independent subsets were suitable for applying the beam search. This simple method ended up being used to create many of the subsets that the ships utilized. Figure~\ref{fig:subset_classical} demonstrates the overlap between these subsets and the final submitted solution, where many more of the green and red markers (asteroids within the solution) are green, indicating that they came from a subset generated in this fashion. Most of these markers exist in a clear region separated via ecliptic longitude. These are the subsets that permit efficient initial and terminal transfers to maximize the possible mining time. However, several ships in the solution did not come from these subsets (seen in red). Instead, these came from a secondary method based on graph theory.

\begin{figure}[hbt!]
    \centering
    \includegraphics[width = 14cm]{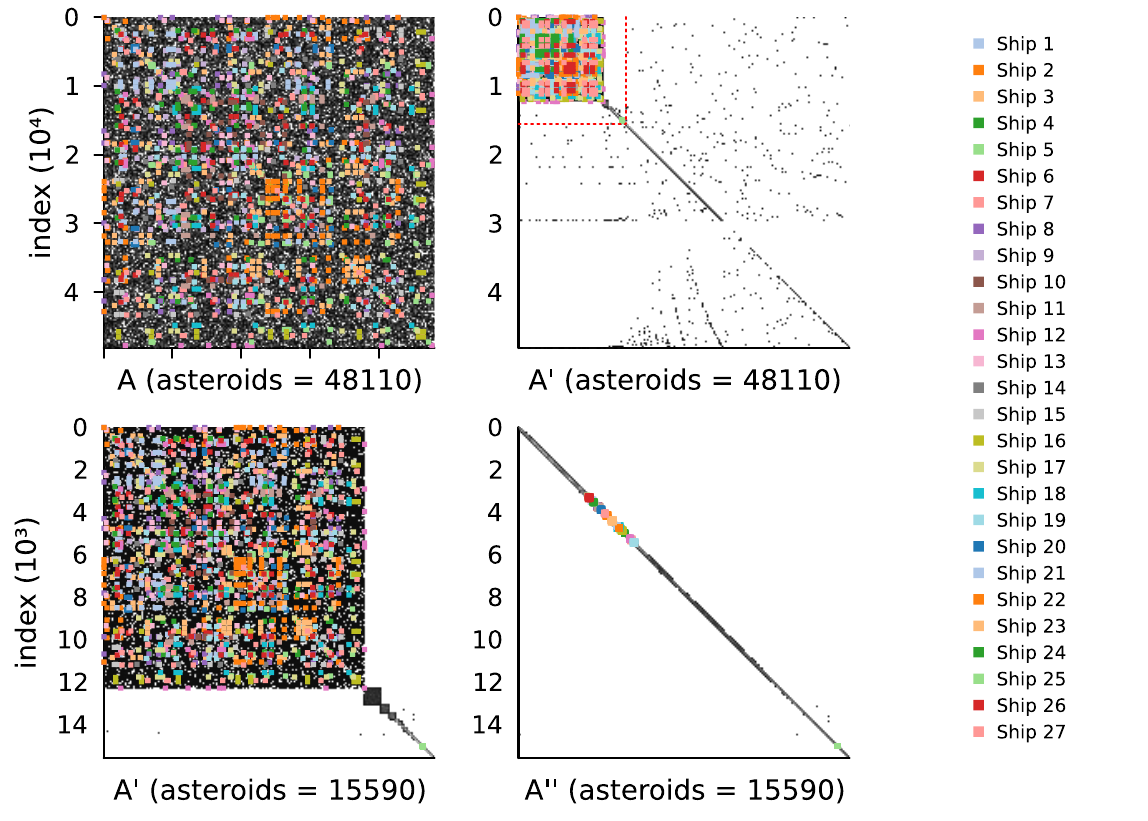}
    \caption{Visual representation of the adjacency matrix $A$ (top left), $A'$ (top right), $A'$ after the filtering (bottom left) and $A''$ (bottom right). Transfers that are used in the final submitted solution are highlighted.}
    \label{fig:clusteringCri}
\end{figure}

To do so, we first organized the data defined by each asteroid's subset as a directed graph \citep{cormen2022introduction}. The adjacency matrix associated with this graph,  $A$, is a sparse matrix of size $N \times N$, where $N$ is the number of asteroids. The generic element, $a_{ij}$, is one if the $j$th asteroid is in the $i$th subset and zero otherwise. A visual representation of $A$ is provided in the top left panel of Figure~\ref{fig:clusteringCri}. Successively, we decomposed the graph into strongly connected components, that is, sets of asteroids that are mutually reachable through a path in the graph. Such strongly connected components define our independent subsets. To visualize them, the rows and the columns of $A$ are topologically reordered such that the asteroids appertaining to the same independent subset are next to each other, and subset size is in descending order. The reordered adjacency matrix, $A'$, has diagonal blocks representing the identified independent subsets, and it is shown in the top right panel of Figure~\ref{fig:clusteringCri}. Knowing from the preliminary analysis that missions with around 10 asteroids would be appropriate, subsets with at least 20 asteroids were considered as input to the beam search to improve the chance of obtaining good asteroid sequences. In this way, the number of asteroids could be reduced from 48,110 to 15,590, divided into 40 independent subsets. The larger cluster includes approximately 80$\%$ of the asteroids.

The selected independent subsets for the beam search are shown in Figure~\ref{fig:clusteringCri}, in the bottom left panel. In the same figure, the asteroids of the submitted solution and all the asteroids connected are highlighted with colors. The asteroids related to the same ship, shown with the same color, are spread and almost exclusively part of the larger cluster in $A'$. That means that the objective of finding independent ships was not achieved since the majority of the ships are part of the same independent subset. Therefore, after the competition, a further assessment was done in an attempt to organize the data better. This reorganization, achieved by the use of the MATLAB function \emph{symrcm} applied to $A'$, results in an almost diagonal adjacency matrix, $A''$, shown in the bottom right panel in Figure~\ref{fig:clusteringCri}. Notably, the asteroids related to the same ship are now close to each other. This means that small and (partially) independent subsets could have been defined by taking portions of $A''$. This step would have further aided the beam search by reducing the search space.

\subsection{\label{sec:BeamSearchAndrea}Beam Search}

For each independent asteroid subset identified in Section~\ref{sec:Subsets}, a selection of mining ship trajectories are generated. These are built such that each ship: (1) departs Earth for the asteroid belt, (2) deploys miners with a minimum number of 6 asteroids, (3) waits at least 1 year at the last asteroid in the chain before picking up mined mass, (4) visits the same asteroids to collect the mined mass (self-cleaning) and (5) returns back to Earth. This structure means self-cleaning trajectories are created and that the ship does not execute a transfer between the deployment and collection stages of the trajectory.

To build such trajectories, an approach based upon the beam search is used, which has emerged as a standard approach to solve GTOC problems \citep{Izzo2016, ZHANG2023819, armellin2022team, bellome2023phd}. In beam search strategies, the construction of mining ship trajectories is performed by adding one asteroid at a time. Two steps are usually needed: (1) branching new trajectory legs and (2) selecting promising options to be kept for further branching. These steps are repeated until a stopping criterion is met, such as when a ship returns to the Earth. In the branching step, one finds the trajectory linking two asteroids and computes the cost between them. Lambert arcs are used as an approximation for the low-thrust trajectories and to estimate the $ \Delta v $ between asteroids. The time-of-flight is discretized within [500, 700] days for the initial and terminal transfers and [170, 250] days for intermediate transfers, both with a 5-day step size. The initial and terminal transfers are further adjusted by the \SI{6}{\km\per\second} provided for injection or flyby in the problem. Then, in the selection step, each trajectory option is ranked with respect to the accumulated $ \Delta v $, and only a limited number of them (the beam width) are kept for further branching. Throughout the GTOC 12 problem, beam widths of 5 to 20 were used, but a beam width of 20 was the most common. Smaller widths were used depending on the subset size and computational factors. Additionally, only asteroid transfers that require an acceleration below the maximum admissible by the thrust limitations are considered \citep{Izzo2016, gtoc12ProblemDescription}.

An important consideration to make with the GTOC 12 problem is the search order of the beam search. A naive approach, which was widely used, is to build a trajectory asteroid by asteroid in the forward direction, starting the beam search at the initial transfer and finishing at the terminal transfer. Alternatively, the initial deployment can be created, followed by the collection leg, but in reverse order. This has the advantage of fitting more naturally with the objective function as more efficient transfers (allowing shorter times-of-flight) nearer to the initial and terminal transfers have the effect of allowing for the movement of more rendezvous times. However, this search order means that the final mass will not be defined when it is needed to select the terminal transfer (and then the remaining collection transfers). It is approximated as 1200 kg, representing a mission that returns 700 kg of mined mass. Out investigations found that this approach tended to be superior for the implementation of a beam search because a beam search is inherently greedy and does not fully iterate the search space. In either case, the sequences were returned to the naive ordering prior to our subsequent steps.

This approach to designing a beam search for spacecraft trajectory optimization problems is typically very computationally efficient, largely due to the low cost of solving the Lambert problems ($ \leq \SI{1}{\micro\second}$ is needed for each Lambert problem to be solved \citep{izzo2015revisiting, russell2022complete}). However, the resulting trajectories might lie far from the corresponding low-thrust solutions, and the difference between these is difficult to know a priori \citep{Mateas2023}. This presented a difficult challenge to solve during the competition: we found that some of the trajectories generated from the beam search using the Lambert $\Delta v$ directly were difficult to make feasible in the low-thrust refinement (explained in Section~\ref{sec:ManualRefinement}) despite the use of margins for the impulsive to low-thrust conversion. As a result, many ships' trajectories either used too much fuel or the time of flight had to be excessively increased. This significantly degraded the performance of many solutions from the Lambert-based beam search. A heuristic constraint was developed to supplement the direct Lambert-based $\Delta v$ selection within the beam search in an attempt to address this problem. The beam search does not consider asteroid-to-asteroid transfers that do not meet the heuristic. This heuristic was generated by a trial and error approach, wherein a limited range of mathematical functions and parameters are tested manually in the beam search to find those that tended to give good results.
\begin{align}
    \text{tof} \geq \left\{\begin{array}{lr}
            c_1 \sqrt[3]{\Delta v} & \text{(initial transfer)}\\
            c_2 \sqrt[3]{\Delta v} & \text{(terminal transfer)}\\
            c_3 \Delta v (1 + m_{\text{current}}/m_{\text{max}}) + h c_4 & \text{(deploying)}\\
            c_5 \Delta v + h c_6 & \text{(collecting)}
        \end{array}\right.
        \label{eq:beam_heuristic}
\end{align}

In Equation~\ref{eq:beam_heuristic} the $\Delta v$ and tof are in \SI{}{\kilo\meter\per\second} and days respectively, and are from the current transfer only (not accumulated). The $c$ parameters are derived, with many having differing units; $c_1$ and $c_2$ are in ($\text{days}\;\text{s}^{1/3}\;\text{km}^{-1/3}$), $c_3$ and $c_5$ are in ($\text{days}\;\text{s}\;\text{km}^{-1}$) and $c_4$ and $c_6$ in $\text{days}$. The $\Delta v$ is approximated by solving Lambert's problem. The $h$ parameter is a scaling factor that takes the value of the index of how far away we are from the initial or terminal transfer. For example, if $h=5$ we are at the 5th deployment after the initial transfer or alternatively at the 5th collection away from the terminal transfer. This is intended to allow for faster, less fuel-efficient transfers at the start or end of the entire chain compared to the middle. By default, the parameters took the values $c=[75, 65, 72, 6, 80, 10]$. These parameters can then be manually tuned based on the characteristics of each subset. 
\begin{table}[hbt!]
    \centering
    \caption{Comparison of low-thrust mining ship trajectories generated from a single subset of 56 asteroids using either the direct Lambert or heuristic-based beam search.}
    \begin{tabular}{cccccc}\toprule
        \multirow{2}{*}{Asteroids} & \multirow{2}{*}{Beam Search} & \multicolumn{2}{c}{Direct Solution} & \multicolumn{2}{c}{Post-competition Refinement} \\ \cmidrule(lr){3-4} \cmidrule(lr){5-6} & & Mined Mass (kg) & Fuel Remaining (kg) & Mined Mass (kg) & Fuel Remaining (kg) \\
        \midrule
        $7$ & lambert & $609.6$ & $-195.4$ & $614.8$ & $3.6$\\
         & heuristic & $571.9$ & $126.2$ & $605.1$ & $2.2$\\
        $8$ & lambert & $632.6$ & $-258.6$ & $620.5$ & $4.0$\\
         & heuristic & $620.5$ & $-50.1$ & $639.3$ & $3.2$\\
        $9$ & lambert & $671.6$ & $-391.2$ & $638.1$ & $1.7$\\
         & heuristic & $653.6$ & $-163.2$ & $627.1$ & $2.2$ \\ 
        \bottomrule
    \end{tabular}
    \addtolength{\tabcolsep}{-5pt} 
    \label{tab:refined_lt_lambert}
\end{table}

After the competition ended, a refinement of the asteroid rendezvous times on some solutions generated with the different beam search implementations was performed. The results are shown in Table~\ref{tab:refined_lt_lambert}. Generally, after refinement, both methods perform similarly in terms of mined mass, perhaps with the direct Lambert approach permitting slightly higher quality solutions.

Both beam search approaches generally result in a negative fuel remaining, especially with increasing amounts of asteroid rendezvous. In the competition, this needed to be corrected during the manual refinement stage as in Section~\ref{sec:ManualRefinement}. This step can be challenging, especially if a large amount of fuel needs to be found. This made the heuristic-based beam search more appealing during the competition as it generally produced solutions that were closer to feasibility with less of a fuel deficit to make up. The heuristic parameters can also be tuned to find solutions with lower mined mass but more fuel remaining. In fact, the introduction of these controls on the beam search through the use of the heuristic was one of the key advancements to our strategy in the last few days of competition, simply because it enabled to creation of solutions that are directly feasible or near-feasible. In all cases, the differences between the direct output of the beam search and the post-competition refinement make it clear that further work is needed to deeply understand the conversion between Lambert arcs and low-thrust arcs.

\section{\label{sec:TrajectoryOpt}Low-thrust Trajectory Optimization}
A mining ship's trajectory is determined by electric propulsion with specific impulse $I_{\text{sp}}=4,000$ s and maximum thrust $T_{\text{max}}=0.6$ N. This clearly places the mining ship spacecraft into the low-thrust regime, and the associated mass-optimal problem can be solved efficiently via direct or indirect methods \citep{indirect,moranteSurveyLowThrustTrajectory2021}. After an initial investigation comparing both approaches, we selected to use direct methods in our final submitted solution procedure. Direct methods offered several advantages over indirect ones, including needing no interpolation in the solution file format and matching the mass-optimal performance whilst permitting faster computational times. 

\subsection{\label{sec:SCP}Sequential Convex Programming}
The use of SCP was fundamental to the competitive performance of direct methods. Applying the general approach of SCP \citep{malyutaConvexOptimizationTrajectory2022}, we start by finding an appropriate linearization of the dynamical system. The dynamics of the mining ship are provided with the state $\mathbf{x}=[\mathbf{r}, \mathbf{v}, m]$ and control $\mathbf{u}=[\mathbf{T}]$:
\begin{align}
    \mathbf{\dot{x}} = f(\mathbf{x}, \mathbf{u}) = \left\{\begin{array}{l}
            \mathbf{\dot{r}} = \mathbf{v} \\
            \mathbf{\dot{v}} = -\frac{\mu}{r^3} \mathbf{r} + \frac{T_{\text{max}}}{m} \mathbf{T}\\
            \dot{m} = -\frac{T_{\text{max}}}{I_{\text{sp}} g_0} T
        \end{array}\right.
\end{align}
where $\mu$ is the gravitational constant of the Sun, $m$ the mass, and $0 \leq T \leq 1$ the thrust of the mining ship, which is normalized to $T_{\text{max}}$. A non-dimensionalization is performed where the length scale is set to $1$ AU, mass scale to $3000$ kg (the maximum permitted launch mass of a mining ship), and $\mu$ to $1$ with the velocity, time, and thrust scaling derived. 

A Lambert arc is used as the reference trajectory for linearization. This was often close to the optimal low-thrust trajectory due to the relatively short time-frames experienced when compared to some low-thrust applications. Single- or multi-revolution Lambert solver variants are used depending on the transfer length, and the transfer with the lowest $\Delta v$ cost was selected. Thus, given a reference trajectory and thrust profile $(\mathbf{\bar{x}}, \mathbf{\bar{u}})$, the dynamical linearization can be performed, which is calculated along segments of the reference trajectory. For the GTOC 12 problem, the number of segments $N$ is selected based on an intended segment timespan across the entire timespan ($t_s, t_e$) of the transfer. For example, a 201-day transfer with 5-day segments would have a total of fourty 5-day segments with a terminal 1-day segment. A zero-order-hold (ZOH) control is assumed on each segment, so the control $\mathbf{u}$ is held piecewise constant. Thus, we obtain the discrete form of the dynamics for $n = 1, 2, ..., N$, which is able to be used as a convex constraint:
\begin{align} \label{eq:dynamics_linear}
    \mathbf{x}_{n+1} = A_n\mathbf{x}_n + B_n\mathbf{u}_n + C_n 
\end{align}
with
\begin{align}
    A_n = \left. \frac{\partial}{\partial \mathbf{x}} \int_{t_n}^{t_{n+1}}\mathbf{\dot{x}}\,\text{d}t\;\right|_{(\mathbf{\bar{x}}_n, \mathbf{\bar{u}}_n)} && B_n = \left. \frac{\partial}{\partial \mathbf{u}} \int_{t_n}^{t_{n+1}} \mathbf{\dot{x}}\,\text{d}t\; \right|_{(\mathbf{\bar{x}}_n, \mathbf{\bar{u}}_n)} && 
    C_n = \mathbf{\bar{x}}_n - A_n \mathbf{\bar{x}_n} - B_n \mathbf{\bar{u}_n}
\end{align}
The matrices $A_n$ and $B_n$ are the state transition matrix (STM) and the input matrix computed via automatic differentiation (AD). The process to use AD is as follows: the dynamical equations are placed into an adaptive step-size numerical integrator, and AD is used to calculate the Jacobian of the output with respect to the input state and control. The Tsit5 \citep{tsitourasRungeKuttaPairs2011} numerical integrator was used from \texttt{DifferentialEquations.jl} \citep{rackauckasDifferentialEquationsJlPerformant2017} with absolute tolerance $10^{-12}$ and relative tolerance $10^{-12}$, and \texttt{ForwardDiff.jl} \citep{revelsForwardModeAutomaticDifferentiation2016} was used to perform the forward-mode automatic differentiation. The dynamics are left in their Cartesian form as they fit naturally within the solution file format and problem specification, and the SCP algorithm converged quickly without many convergence problems. Alternative coordinate representations are likely to have shown superior linearization and convergence \citep{hofmannComputationalGuidanceLowThrust2023a} but would have required additional complexity for only a small payoff during the competition time-frame, considering the speed of SCP with Cartesian elements.

We also utilized an alternative expression of the dynamics based on a logged-mass formulation \citep{acikmeseConvexProgrammingApproach2007}, which tended to exhibit superior convergence characteristics when compared to the true-mass formulation. The change of variables is as follows:
\begin{align}
    \mathbf{\Gamma} = \frac{\mathbf{T}}{m} && w = \ln m
\end{align}
leaving the dynamical equations with the adjusted state $\mathbf{x}=[\mathbf{r}, \mathbf{v}, w]$ and control $\mathbf{u}=[\mathbf{\Gamma}]$:
\begin{align}
    \mathbf{\dot{x}} = f(\mathbf{x}, \mathbf{u}) = \left\{\begin{array}{l}
            \mathbf{\dot{r}} = \mathbf{v} \\
            \mathbf{\dot{v}} = -\frac{\mu}{r^3} \mathbf{r} + T_{\text{max}} \mathbf{\Gamma} \\
            \dot{w} = -\frac{T_{\text{max}}}{I_{\text{sp}} g_0} \Gamma
        \end{array}\right.
\end{align}
This is converted into a convex constraint using the same process as the true-mass formulation. We found that this formulation had problems passing the solution verification process. This is due to the slight differences in how the ZOH control is interpreted between the two formulations. The thrust is kept constant throughout a segment in the true-mass formulation but not in the logged-mass formulation. Rather, $\Gamma$ is kept constant, so the thrust decreases with the mass. Two solutions to this problem came to mind during the competition timeframe: either decreasing the segmentation time significantly when writing the solution files (so that the thrust decrease would be minimal) or running a secondary SCP with the true-mass formulation using a reference solution from the logged-mass formulation. We elected to use the latter as the logged-mass reference solution allowed for very fast (often single iteration) convergence of the secondary SCP.

Most of the remaining constraints are able to be put into a lossless convex form. Firstly, for the control magnitude constraint, we use a convex second-order-cone (SOC) to obtain the $L_2$ norm of the control vector:
\begin{align} \label{eq:control_mag}
    u_n \geq \|\mathbf{u}_n\|_2
\end{align}
This constraint will bind to equality at optimality because the objective function indirectly minimizes the control. We can then enforce the control limits based on the mass formulation used:
\begin{align} \label{eq:control_limit}
    u_n \leq 1 \quad (\text{true-mass}) && u_n \leq e^{-\bar{w}_n} (1 - w_n + \bar{w}_n) \quad(\text{logged-mass})
\end{align}
The true-mass problem has a simple form for this constraint due to the normalization procedure, but the logged-mass variant needs an additional linearization step. We take the Taylor expansion of $\Gamma = e^{-w} T$ around a reference $\bar{w}$ to obtain the appropriate control limit, which is updated at each iteration of the SCP. The initial and terminal location constraints are:
\begin{align} \label{eq:initial_state}
    \left[\begin{array}{c}
        \mathbf{r}_1 \\ \mathbf{v}_1
    \end{array} \right] = \left[\begin{array}{c}
        \mathbf{r}_\text{d} \\ \mathbf{v}_\text{d} + \Delta \mathbf{v}_\text{d}
    \end{array} \right] \\ \label{eq:final_state}
    \left[\begin{array}{c}
        \mathbf{r}_{N+1} \\ \mathbf{v}_{N+1}
    \end{array} \right] = \left[\begin{array}{c}
        \mathbf{r}_\text{a} \\ \mathbf{v}_\text{a} + \Delta \mathbf{v}_\text{a}
    \end{array} \right]
\end{align}
An additional $\Delta v$ is added to the velocity of each of these. This is to allow for injection or flyby $\Delta v$ to be optimized simultaneously with the low-thrust control, which was important to the GTOC 12 problem because an additional $6$ km/s was permitted for these with Earth. We also need to apply control limits applied to these, again using SOC constraints:
\begin{align} \label{eq:dv_start_limit}
    \Delta v_\text{d, max} \geq \|\Delta \mathbf{v}_\text{d}\|_2 \\ \label{eq:dv_end_limit}
    \Delta v_\text{a, max} \geq \|\Delta \mathbf{v}_\text{a}\|_2
\end{align}
Finally, we apply the initial mass constraint:
\begin{align} \label{eq:mass_start}
    m_1 = m_\text{d} \quad (\text{true-mass})&& w_1 = w_\text{d} \quad(\text{logged-mass})
\end{align}
This was left as an equality constraint, but it can be relaxed to inequality if required. During the GTOC 12 time frame, this was done to quantify the level of infeasibility of certain trajectories, particularly the initial transfer from Earth to asteroids. Acceleration increases when the initial mass decreases, making the trajectory more easily feasible. Combining all these constraints with the objective of maximizing the final mass, the entire optimization problem is therefore:
\begin{mini}
    {\mathbf{u}, \Delta \mathbf{v}_\text{d}, \Delta \mathbf{v}_\text{a}}{
    \left\{\begin{array}{l}
         -m_{N+1}\quad (\text{true-mass}) \\
         -w_{N+1}\quad (\text{logged-mass})
    \end{array}\right.}{}{}
    \addConstraint{\eqref{eq:dynamics_linear}, \eqref{eq:control_mag}, \eqref{eq:control_limit}}{}{n=1,2,\ldots,N}
    \addConstraint{\eqref{eq:initial_state}, \eqref{eq:final_state}, \eqref{eq:dv_start_limit}, \eqref{eq:dv_end_limit}, \eqref{eq:mass_start}}{}
    \label{eq:convex_problem}
\end{mini}
The SCP process repeatedly solves \eqref{eq:convex_problem} using a convex optimizer and updates the linearized constraints \eqref{eq:dynamics_linear} and \eqref{eq:control_limit} with the optimal solution from the previous iteration. The convergence of the algorithm is determined by the accuracy of the linearization when compared to truth. It is important to note that no trust regions, soft relaxations, or virtual controls are used to aid the convergence of the formulation.

Following the development of the SCP that can solve a single leg of a journey, the next step was to assemble these to optimize the entire trajectory for a single ship. The intent is to have a single method that can go from a given ID and time list for each ship to an optimal control profile that can be validated. 

Two different methodologies are developed. The first is to solve a trajectory sequentially (`sequential method') by solving each leg of the journey in order, ensuring that the final mass of each leg is the initial mass for the next after any changes due to drop-off or pickup. Depending on the journey length, approximately 20 SCP problems are solved to optimize an entire trajectory, but these problems are kept relatively small in size. 

Alternatively, a single SCP problem can be assembled, encompassing the whole trajectory (`unified method'). All of the variables and constraints from each leg of the journey are added to the same SCP problem, with the addition of a single convex constraint linking the mass between subsequent legs. The initial mass of the first leg is fixed, but for all intermediate legs, a constraint is added so that the initial mass is equal to the terminal mass of the previous leg after any changes due to drop-off or pickup. The objective is to maximize the terminal mass of the final leg. Also, this formulation allows for the initial mass constraint to be relaxed, so the launch mass can be reduced from the maximum if it would be optimal to do so.
\begin{table}[hbt!]
    \centering
    \caption{Details on the $9$ asteroid self-cleaning case used to demonstrate the SCP solve process, which returns $711.29\text{kg}$ of mined mass to Earth.}
    \begin{tabular}{p{0.2\linewidth} p{0.75\linewidth}}
        \hline
        \\[-3pt]
        ID Sequence & [0, 25279, 27464, 24025, 37117, 3750, 3196, 51591, 47047, 14516, 14516, 3750, 47047, 37117, 3196, 51591, 27464, 25279, 24025, -3] \\
        Time Sequence (MJD) & [64458.0, 64983.0, 65148.0, 65388.0, 65643.0, 65848.0, 66043.0, 66258.0, 66588.0, 66773.0, 68023.0, 68188.0, 68448.0, 68648.0, 68808.0, 68908.0, 69083.0, 69193.0, 69353.0, 69807.0] \\ 
        \\[-3pt]
        \hline
    \end{tabular}
    \label{tab:example_for_convex}
\end{table}
\begin{table}[hbt!]
    \centering
    \addtolength{\tabcolsep}{5pt} 
    \caption{Performance characteristics for both the unified and sequential formulations finding the optimal control for the self-cleaning case in Table~\ref{tab:example_for_convex} across a range of leg segmentations. Times are averaged over a sample size of $25$ and calculated with an Intel Core i7-12700 processor. Total time is the entire time of the whole algorithm, not just including the AD and Solve portions.}
    \begin{tabular}{llrccc}\toprule
        \multirow{2}{*}{Segmentation (days)} & \multirow{2}{*}{Formulation} & \multirow{2}{*}{Fuel Remaining (kg)} & \multicolumn{3}{c}{Time (s)} \\ \cmidrule(lr){4-6} & & & AD & Solve & Total \\ 
        \midrule
        $1$ & sequential & $4.389$ & $0.544$ & $1.509$ & 2.105 \\
         & unified & $4.429$ & $0.609$ & $4.007$ & 4.661 \\
        $2$ & sequential & $4.144$ & $0.317$ & $0.684$ & 1.046 \\
         & unified & $4.186$ & $0.359$ & $1.532$ & 1.920 \\
        $5$ & sequential & $3.013$ & $0.137$ & $0.303$ & 0.469 \\
         & unified & $3.220$ & $0.185$ & $0.615$ & 0.816 \\
        $10$ & sequential & $0.130$ & $0.095$ & $0.175$ & 0.294 \\
         & unified & $0.347$ & $0.111$ & $0.304$ & 0.428 \\
        $20$ & sequential & $-10.692$ & $0.073$ & $0.102$ & 0.198 \\
         & unified & $-10.451$ & $0.078$ & $0.158$ & 0.247 \\
        $50$ & sequential & $-96.668$ & $0.053$ & $0.059$ & 0.131 \\
         & unified & $-96.376$ & $0.047$ & $0.093$ & 0.148 \\
        \bottomrule
    \end{tabular}
    \addtolength{\tabcolsep}{-5pt} 
    \label{tab:convex_performance}
\end{table}
The unified method tended to exhibit slightly superior optimality when compared to the sequential method, which can be seen across the statistics presented in Table~\ref{tab:convex_performance}. However, the unified method tended to be approximately $2$x slower. Both methods were used throughout the GTOC 12 time frame, but we used the sequential method within the final submitted solution.
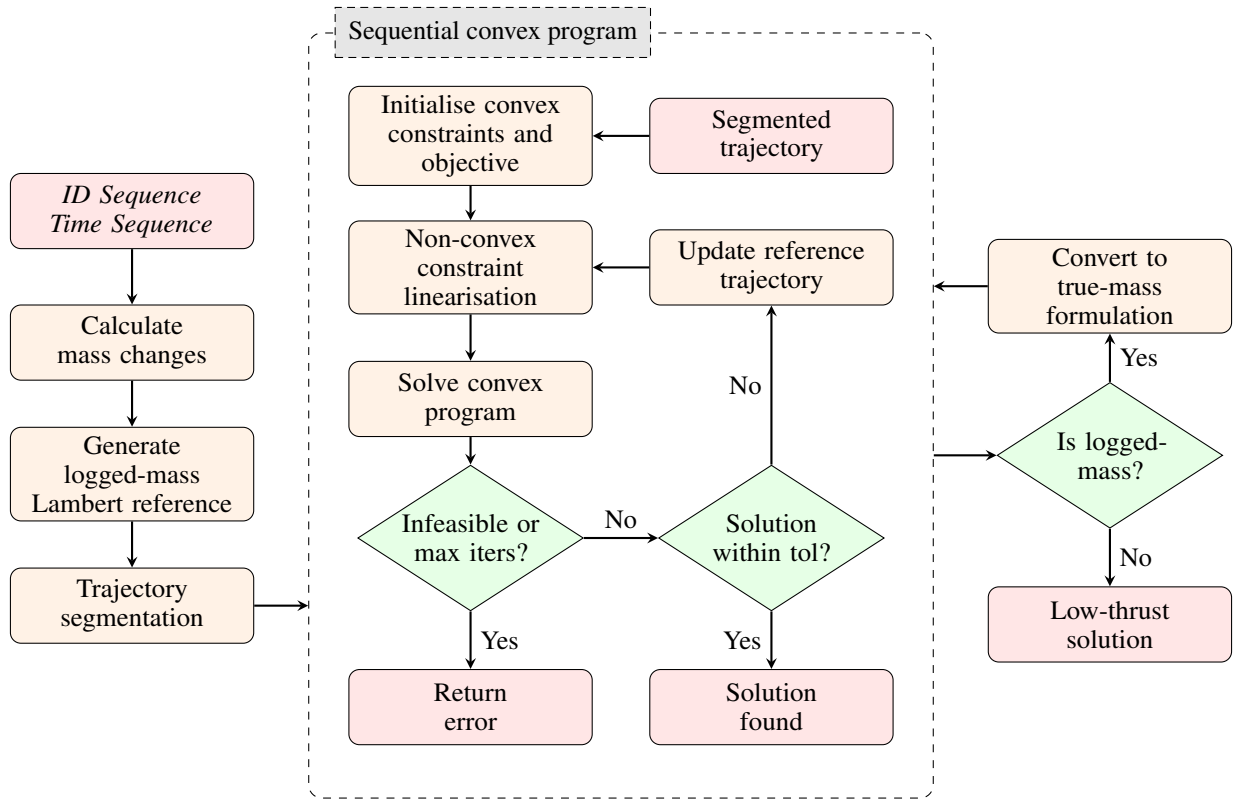
\begin{figure}[hbt!]
    \centering
    \begin{tikzpicture}[node distance=1.75cm]
    \node (convex_input) [startstop] {Segmented\\trajectory};
    \node (convex_init) [process, left of=convex_input, xshift=-2.25cm] {Initialise convex\\constraints and\\objective};
    \node (reference_update) [process, below of=convex_input] {Update reference\\trajectory};
    \node (linearisation) [process, below of=convex_init] {Non-convex\\constraint\\linearisation};
    \node (convex_solving) [process, below of=linearisation] {Solve convex\\program};
    \node (convergence) [decision, below of=convex_solving, yshift=-0.1cm] {Infeasible or\\max iters?};
    \node (convergence2) [decision, right of=convergence, xshift=2.25cm] {Solution\\within tol?};
    \node (failed_convex) [startstop, below of=convergence, yshift=-0.5cm] {Return\\error};
    \node (successful_convex) [startstop, below of=convergence2, yshift=-0.5cm] {Solution\\found};

    \node[fit=(convex_init)(linearisation)(convex_solving)(convergence)(convex_input)(failed_convex)(successful_convex),myfit] (myfit1) {};
    \node[mytitle] at (myfit1.north west) {Sequential convex program};

    \node (input_sequence) [startstop, left of=convex_input, xshift=-6.75cm, yshift=-1.0cm] {\textit{ID Sequence}\\\textit{Time Sequence}};
    \node (mass_sequence) [process, below of=input_sequence] {Calculate\\mass changes};
    \node (lambert_reference) [process, below of=mass_sequence] {Generate\\logged-mass\\Lambert reference};
    \node (segmentation) [process, below of=lambert_reference] {Trajectory\\segmentation};

    \node (logged_conversion) [process, right of=convex_input, xshift=2.75cm, yshift=-2.0cm] {Convert to\\true-mass\\formulation};
    \node (logged_mass_check) [decision, below of=logged_conversion, yshift=-0.5cm] {Is logged-\\mass?};
    \node (final_output) [startstop, below of=logged_mass_check, yshift=-0.5cm] {Low-thrust\\solution};

    \draw [arrow] (convex_input) -- (convex_init);
    \draw [arrow] (convex_init.south) -- (linearisation.north);
    \draw [arrow] (linearisation) -- (convex_solving);
    \draw [arrow] (convex_solving) -- (convergence);
    \draw [arrow] (convergence) -- node[midway,above] {No} (convergence2);
    \draw [arrow] (convergence2.north) -- node[midway, left] {No}(reference_update);
    \draw [arrow] (reference_update) -- (linearisation);
    \draw [arrow] (convergence.south) -- node[midway,right] {Yes}(failed_convex);
    \draw [arrow] (convergence2.south) -- node[midway, left] {Yes}(successful_convex);

    \draw [arrow] (input_sequence) -- (mass_sequence);
    \draw [arrow] (mass_sequence) -- (lambert_reference);
    \draw [arrow] (lambert_reference) -- (segmentation);
    \draw [arrow] (segmentation.east) -- (segmentation.east-|myfit1.west);

    \draw [arrow] (myfit1.east|-logged_mass_check.west) -- (logged_mass_check.west);
    \draw [arrow] (logged_mass_check) -- node[midway,right] {Yes}(logged_conversion);
    \draw [arrow] (logged_mass_check) -- node[midway,right] {No}(final_output);
    \draw [arrow] (logged_conversion.west) -- (logged_conversion.west-|myfit1.east);
    
    \end{tikzpicture}
    \caption{Schematic of the algorithm for applying SCP to the GTOC 12 problem.}
    \label{fig:scp_process}
\end{figure}

Our solution process, seen in Figure~\ref{fig:scp_process}, consisted of solving the SCP with the sequential method and logged-mass formulation for each mining ship trajectory, with a convergence parameter of $10^{-5}$, followed by using the output to solve the same problem in the true-mass formulation with a convergence parameter of $10^{-7}$. In terms of implementation, \texttt{JuMP.jl} \citep{lubinJuMPRecentImprovements2023} was used to create and modify the convex problems and MOSEK to solve them. This process provided an output control profile that meets the target error requirements for validation.

\subsection{\label{sec:ManualRefinement}Manual Rendezvous Time Refinement}
The solutions obtained from the beam search are often suboptimal and can be improved significantly by altering rendezvous times. They also may not have enough fuel or even have excess fuel at the final Earth flyby, so it is apparent changes can be made to improve the quality of the solutions.
\begin{figure}[hbt!]
    \centering
    \includegraphics[width = 16cm]{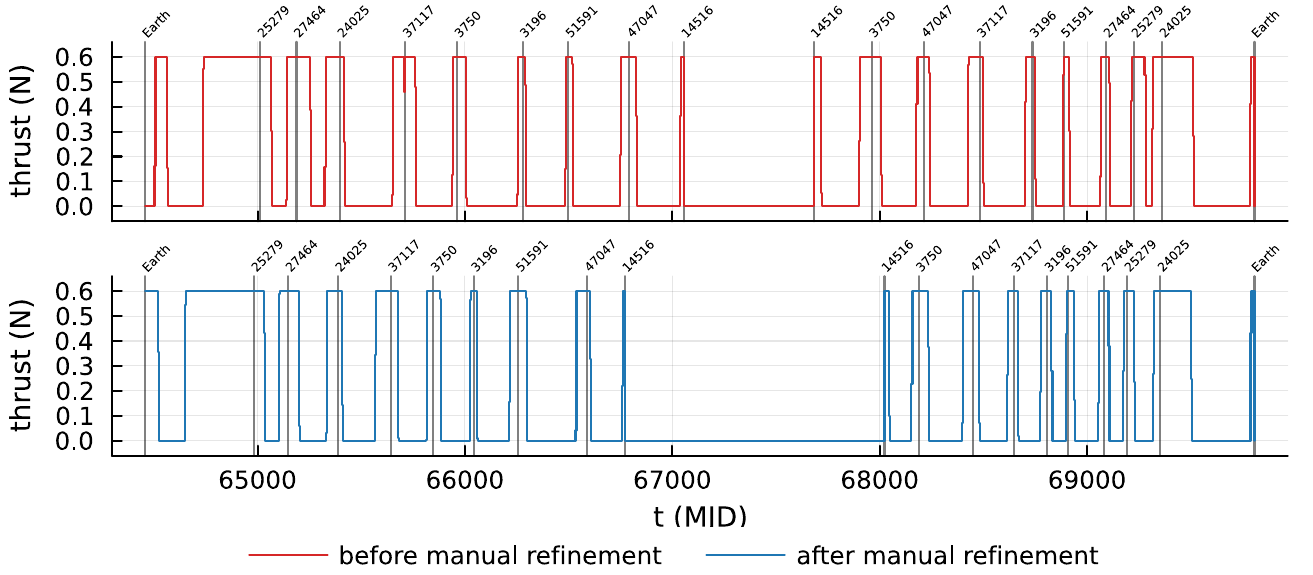}
    \caption{Comparison of the thrust profile for the solution presented in Table~\ref{tab:example_for_convex} before and after manual rendezvous time refinement. The mass returned increases from $649.99\text{kg}$ to $711.29\text{kg}$.}
    \label{fig:manual_ref}
\end{figure}

\begin{figure}[hbt!]
    \centering
    \includegraphics[width = 16cm]{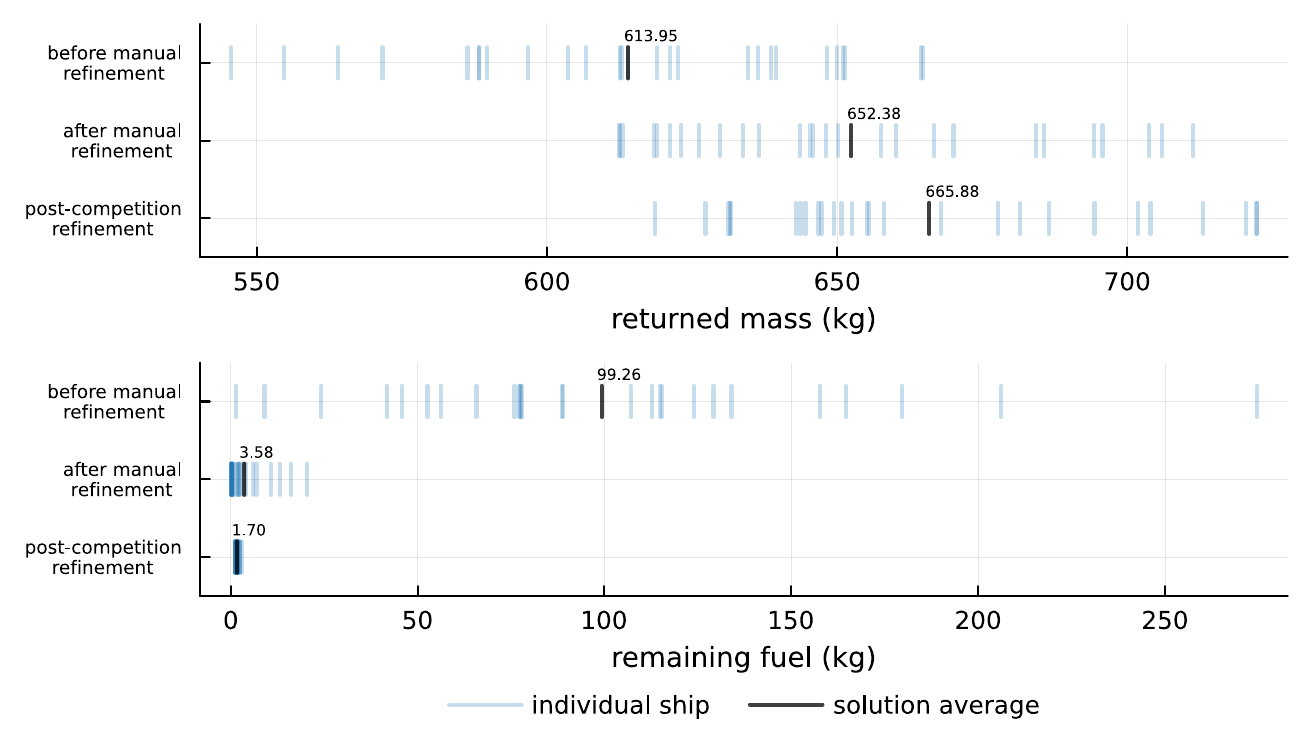}
    \caption{Comparison between the different rendezvous time refinement methods over the $27$ asteroid sequences of the ships in the final submitted solution.}
    \label{fig:ref_compare}
\end{figure}

Due to time constraints, we manually tuned the times of the rendezvous with the intention of maximizing the mass returned by using up all of the fuel. This manual tuning was performed using the output from the low-thrust trajectory optimization and visualizations, an example of which can be seen in both subplots of Figure~\ref{fig:manual_ref}. An iterative process was used where once a change was made to the rendezvous times, the low-thrust trajectory was resolved in order to quantify the change. The strategy was twofold. First, refine the Earth departure and arrival times to optimize the initial and terminal transfer legs with respect to the low-thrust cost and time. Secondly, attempt to shift the deployment times earlier and the collection times later to maximize the returned mass. The time-of-flight of asteroid transfers closer to the initial and terminal transfers had more emphasis placed as a reduction would permit the movement of all the more intermediate transfers (resulting in a larger impact on collected mass).  

Figure~\ref{fig:manual_ref} shows the comparison of the thrust profile between an unrefined $649.99$ kg mission in the top subplot followed by the same mission post-refinement in the bottom subplot returning $711.29$ kg. The unrefined mission was generated using the heuristic beam search on one of the subsets generated. It is clear that the deployer rendezvous times have been substantially shifted earlier and the collection rendezvous times later, allowing for the increase in returned mass. 

Figure~\ref{fig:ref_compare} shows a comparison of the average returned mass and the fuel remaining for the $27$ missions in our final submitted solution using different refinement methods. Before manual refinement, there is a wide spread in the remaining fuel for each mission, with an average of $99.26$ kg. Manual refinement reduced this to $3.58$ kg per ship. Making use of all onboard fuel helps to increase the average mass and, hence, the final score of the solution. Before manual refinement, the returned mass ranges from $546$ kg to $665$ kg. After manual refinement, the average returned mass increased from $613.95$ to $652.38$ kg, the equivalent of $4$ additional ships. When translated to a total score, it allows $3$ additional ships, with the overall returned mass increasing from $14,912.44$ to $17,614.32$. Post-competition refinement shows the same solution can reach $665.88$ kg average, which would enable another additional ship.

\subsection{\label{sec:AttemptedMixing}Mixed Attempts}

The process described thus far is capable of generating self-cleaning missions, i.e., ships that visit each asteroid twice, initially to deploy a miner and later to collect the mined mass. An advantage of constructing solutions in this fashion is the ability to treat ships independently of one another and to reduce the search space complexity during the collection phase of each mission. However, we also invested some time in mixed-ship solutions. These are solutions that require more than one ship to deploy and collect from a given set of asteroids, where each individual ship works in tandem with a set number of other ships to deploy and collect from a larger set of asteroids.

An initial approach was to generate ships that only deployed miners to asteroids and another set of ships that collected mined mass from asteroids. These missions were created using the beam-search approach described in Section~\ref{sec:BeamSearchAndrea}. These deployer and collector missions would then be combined in an optimization process to create Earth-asteroid-Earth missions. This approach was halted by the difficulty of converting these mixed deployer-collector missions into feasible low-thrust trajectories. An interesting question that arose during these investigations was the optimal number of ships in a mixed solution. The extremal options are $N$ totally independent self-cleaning ships on one side and $N$ totally dependent mixed ships on the other. An intermediate possibility is to have several groups of mixed ships that are dependent on each other within a given group but independent of other groups. In hindsight, our approach of combining deployer and collector missions might have started out too ambitious in trying to create very large mixed groups, and this complexity led to us avoiding the difficulties of a fully mixed design.

As an alternative, we considered combining self-cleaning solutions into mixed-ship solutions by adapting the beam-search approach described in Section~\ref{sec:BeamSearchAndrea}. This was done under the assumption that swapping certain asteroids between neighboring self-cleaning ships could remove a constraint on the collection phase and increase the collected mass and hence optimality of the solutions. In the remainder of this section, we describe a $3$-ship, $21$ asteroid solution totaling $1754.69$ kg returned with an average of $584.9$ kg per ship solution. While this approach may generate better mixed-ship solutions, we prioritized the self-clean approach as a means of generating the best solution possible in the remaining competition time frame.

\begin{table}[hbt!]
    \centering
    \caption{Comparison of a $3$-ship self-clean and $3$-ship mixed solution using the same set of $21$ asteroids.}
    \adjustbox{max width=\textwidth}{
    \begin{tabular}{ccccccccc}
        \toprule
        & \multicolumn{2}{c}{Self-cleaning} & \multicolumn{2}{c}{Self-cleaning with 
        post-} & \multicolumn{2}{c}{Mixed} & \multicolumn{2}{c}{Mixed with post-}\\
        & & & \multicolumn{2}{c}{competition refinement} & & & \multicolumn{2}{c}{competition refinement}\\
        \cmidrule(lr){2-3}\cmidrule(lr){4-5}\cmidrule(lr){6-7}\cmidrule(lr){8-9}
        Ship & Mined (kg) & Fuel Remaining (kg) & Mined (kg) & Fuel Remaining (kg) & Mined (kg) & Fuel Remaining (kg) & Mined (kg) & Fuel Remaining (kg) \\
        \midrule
        1 & 554.6 & -55.2 & 552.2 & 1.5 & 625.9 & 42.0 & 639.6 & 1.7 \\
        2 & 556.5 & -66.4 & 553.7 & 2.2 & 596.7 & 15.0 & 622.0 & 2.5 \\
        3 & 609.2 & 14.4 & 623.2 & 2.3 & 532.1 & 15.9 & 549.6 & 1.6\\
        \midrule
        Average & 573.4 & -35.8 & 576.4 & 2.0 & 584.9 & 24.3 & 603.7 & 2.0 \\

        \bottomrule
    \end{tabular}
    }
    \addtolength{\tabcolsep}{-5pt} 
    \label{tab:mixedships}
\end{table}

Table~\ref{tab:mixedships} shows a comparison for a $3$-ship self-cleaning and $3$-ship mixed solution using the same set of $21$ asteroids. The process to obtain the mixed-ship solution was as follows. First, a subset of $25$ asteroids was chosen, and $3$ independent self-cleaning solutions were obtained using the beam-search approach described in Section~\ref{sec:BeamSearchAndrea}. This was done sequentially, where the first self-cleaning solution was obtained, and then the asteroids were removed from the subset before the next self-cleaning solution was found. This is evidently sub-optimal. Once completed, a second round of the heuristic beam search was performed on the $21$ asteroid set for the deploying asteroids and collection asteroids separately. Only two constraints were applied: the first asteroid to be visited by each ship was the same as the self-cleaning solutions, and the first asteroid to be collected was fixed to the last deployed asteroid for each ship. By doing the deployment and collection separately, the times of deployment are fixed across the $3$ ships, and hence, the mined mass could be appropriately calculated during the collection beam search. The main challenge was writing a new function that could check the asteroid visit times across the whole set of ships and not just the current one.

Using this approach, a $3$-ship mixed solution, which is unfeasible due to the final mass, is made feasible, increasing the average mass from $573.4$ kg per ship to $584.9$ kg per ship. This can then be manually refined up to $591.3$ kg per ship or automatically refined post-competition using the same asteroid ordering to $603.7$ kg per ship. Using the same post-competition refinement, the $3$ self-cleaning missions can only reach $576.4$ kg per ship. However, the main limitation remains the sequential nature of the approach, where the mass of each ship gets progressively worse as the heuristic beam search leaves the least desirable asteroids in the subset until last. Post-competition investigations, where the asteroid order is modified, have led to a $637$ kg average solution for the same set of $21$ asteroids. Time constraints prevented us from developing a less greedy beam search better suited to the mixing approach during the competition. However, this mixed solution, along with another $2$-ship solution averaging $574.7$ kg, was good enough to be included in our overall solution as late as $1.5$ days before the final submission.

\section{\label{sec:FinalResult}Mission Selection and Final Result} \label{sec_results}
\subsection{\label{sec:PoolSolutions}Creating a Pool of Solutions}
A pool consisting of 193 self-cleaning solutions, a 3-ship mixed solution, and a 2-ship mixed solution was used to generate the submitted solution. This was created by taking each different asteroid subset generated in Section \ref{sec:Subsets} and running it through the beam search in Section \ref{sec:BeamSearchAndrea} starting from five-asteroid sequences, which are increased in length until the performance is deemed satisfactory (trajectory is feasible, meets constraints and the returned mass is high). This is followed by the low thrust trajectory optimization in Section \ref{sec:TrajectoryOpt} (and the mixing process in Section \ref{sec:AttemptedMixing} for mixed solutions). All feasible solutions are collated to generate a sizable pool of feasible mining ship solutions.

\subsection{\label{sec:Optimiser}Optimal Solution Set Selection}

\begin{figure}[hbt!]
    \centering
    \includegraphics[width = 16cm]{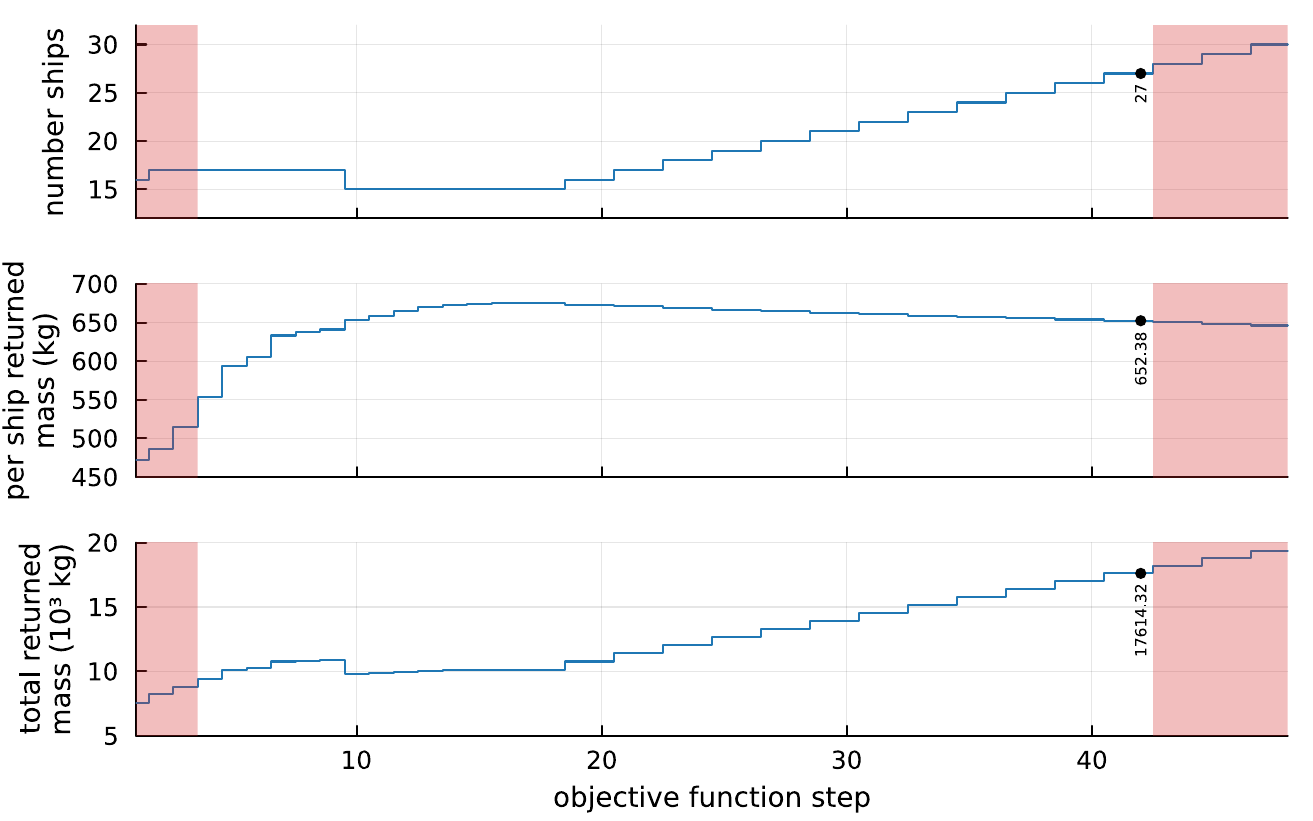}
    \caption{Number of ships, per ship returned mass, and total returned mass as a function of the objective function evaluation step. The red regions represent areas where the combined solution is infeasible with the per-ship mass limit, and the black dot represents the combined solution maximizing returned mass.}
    \label{fig:objectiveIterationCri}
\end{figure}

The methodology to combine the mining ship solutions is based on a random search combined with a deterministic algorithm. The random search defines an initial set of ships to initialize the deterministic algorithm, which completes the solution. Based on this solution, the objective function, the mined mass, is evaluated and maximized using a classical genetic optimizer. The main steps of the optimization process and objective function evaluation are described in Algorithm~\ref{alg:SolutionMixMix}. Two main steps are at the basis of the deterministic algorithm. The first, Step~\ref{alg:AddShip}, ranks all the available ships based on their mined mass and adds the best one that does not conflict (with regards to the asteroids visited) to the current pool of selected ships. Secondly, Step~\ref{alg:SwapShip} performs a ship swap where the ship with the worst mined mass is swapped with the ship with the best mined mass that does not conflict with the current solution. Note that the process described never applies the bonus coefficients $\beta$. This is introduced when the process is rerun via a modification of Step~\ref{alg:SwapShip} so that the ranking function is altered once at the optimal solution. Additionally, to work with mixed ship solutions, these operations are modified to work with sets of ships so that mixed ships are all added and removed from the selected solution pool simultaneously. During the competition, we tested several sets of different optimization approaches, and this tended to work best. The submitted solution was obtained assuming a maximum number of ships of 30, where half of them are initialized randomly. 

Figure~\ref{fig:objectiveIterationCri} demonstrates how the complete solution is built step by step within the objective function. The number of ships (left panel) and the mass and the average mass per ship (right panel) are shown as a function of the step. Note that in the pool of ships, there are mixed solutions, described in Section~\ref{sec:AttemptedMixing}, the reason for which the average mass was chosen to compare and rank the ships instead of the total mass. The 3-ship mixed solution, which is part of the initial pool of ships selected by the optimizer, is swapped on the 8th objective function step for a single ship. This swap implies a decrease in the total mass but an increase in the average mass per ship. A more sophisticated algorithm could have been implemented to swap this mixed solution only if its average mass was smaller than that of the three best ships.
\begin{algorithm}
\caption{Selection of the optimal solution set}\label{alg:SolutionMixMix}
\begin{algorithmic}[1]
\State Set the starting number of selected ships, $n$, and maximum number of ships, $N$
\State Compute set $s$ of $n$ random non-conflicting ships from the pool of solutions
\While{$n \leq N$}
    \State Rank the remaining pool of solutions and add the best non-conflicting ship to $s$ \label{alg:AddShip}
    \While {worst ship in $s$ $<$ best non-conflicting ship}
        \State Swap the worst ship in $s$ with the best available non-conflicting ship \label{alg:SwapShip}
    \EndWhile
\EndWhile
\State Evaluate the best solution set based on the last solution that meets the returned mass constraint
\State Return selected ships, $s$
\end{algorithmic}
\end{algorithm}

The algorithm described is a modified version of that developed in the first weeks of the competition, where the idea was to combine sequences of deploying miners with sequences of collecting the mined mass optimally. So, although unnecessarily convoluted for mixing complete ships, the methodology overall worked well. After the competition, it became apparent that a mixed integer programming (MIP) approach would provide superior computational performance with the same results as the submitted solution. 

\subsection{\label{sec:Performance}Computational Performance}
The time constraints of the GTOC 12 problem made it important to consider the computation times for each part of the solution approach. An approximate overview of the computational performance for the main parts of the solution approach is provided in Table~\ref{tab:performance}. It is important to note that this table only considers the computational times and not the manual time put in by team members. In some cases, the differences between these can be large; for example, the manual refinement stage has a very small computational time when compared to the manual time put in by team members.

\begin{table}[hbt!]
    \centering
    \addtolength{\tabcolsep}{5pt} 
    \caption{Approximate computational overview for the main parts of the selected solution approach. Calculations are performed using an Intel Core i7-12700 processor.}
    \begin{tabular}{lrlrr}\toprule
        Section & Amount (x) & Task & Single Time (s) & Total Time (s)\\
        \midrule
        \ref{sec:ObjectiveFunction} & 1 & Objective function analysis & 60 & 60\\
        \ref{sec:Subsets} & 1 & Lambert transfer precalculation & 3,600 & 3,600\\
        & 1 & Subset calculation \& merging & 600 & 600\\
        \ref{sec:BeamSearchAndrea} & $>$200 & Restricted beam search in subset & 180 & $>$36,000\\
        \ref{sec:SCP} & $>$200 & Mass-optimal direct solution & 1 & $>$200\\
        \ref{sec:ManualRefinement} & $>$50 & Manual refinement of solution & 10 & $>$500\\
        \ref{sec:Optimiser} & 1 & Optimal solution set selection & 600 & 600\\

        \bottomrule
    \end{tabular}
    \addtolength{\tabcolsep}{-5pt} 
    \label{tab:performance}
\end{table}

\subsection{\label{sec:FResult}Final Result}
\begin{figure}[hbt!]
    \centering
    \includegraphics[width = 16cm]{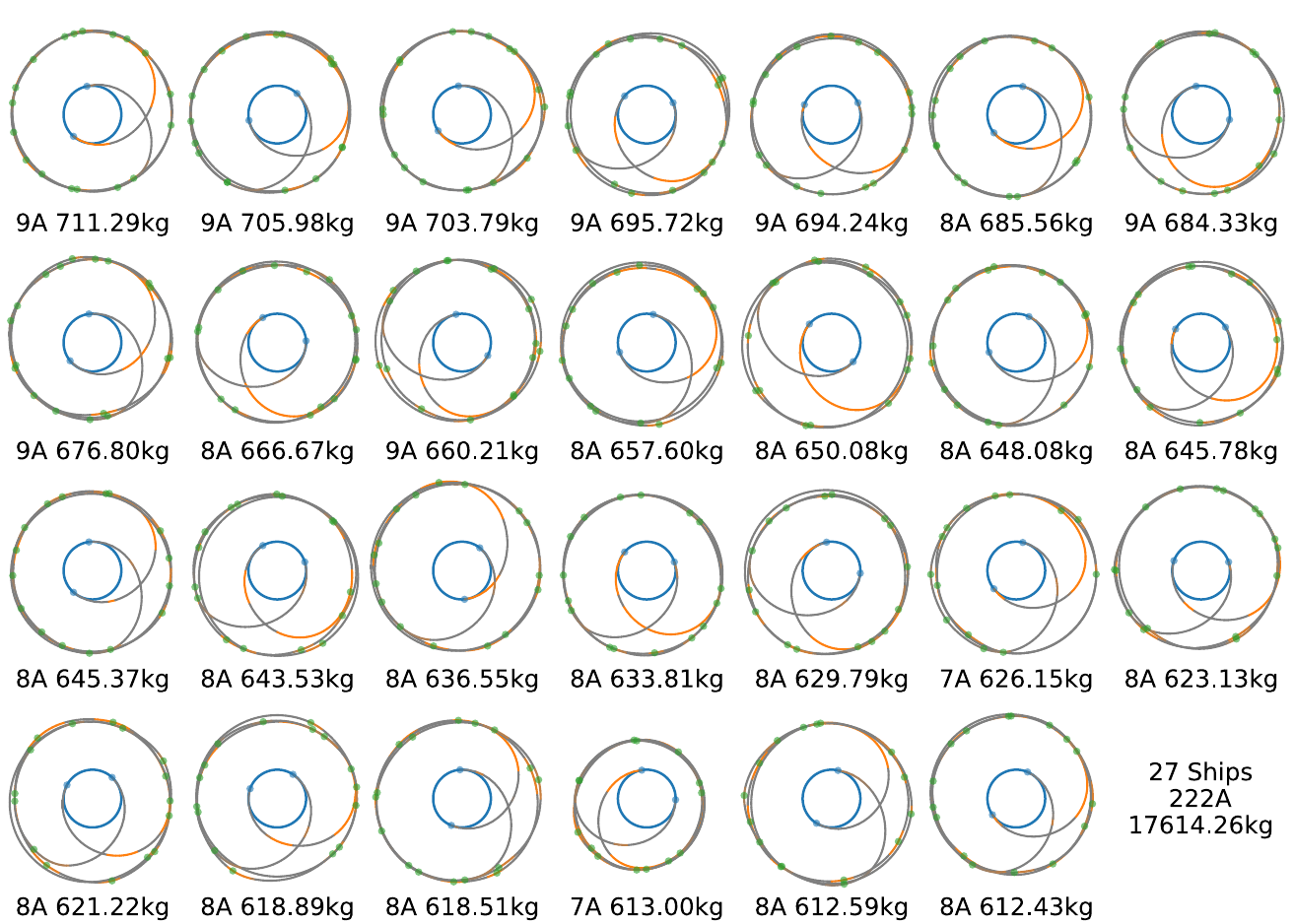}
    \caption{Mining ship trajectories for each of the $27$ ships of the final submitted solution. Ships are ordered in terms of their returned mass. The number of asteroids (denoted by `A' in this figure) and mass are provided underneath each ship's trajectory. Asteroid rendezvous are shown with a green marker, Earth rendezvous with a blue marker, thrust arcs in orange, and coasting arcs in grey.}
    \label{fig:solution_final}
\end{figure}
Figure~\ref{fig:solution_final} provides an overview of all of the component self-cleaning asteroid chains within our team's final submitted solution. The authors have also made a showcase video\footnote{\href{https://www.youtube.com/watch?v=Eb7Br6Zp6ko}{YouTube: Sustainable Asteroid Mining (TheAntipodes GTOC12 Solution)}}. This solution was the result of applying the selection algorithm described in Section~\ref{sec:Optimiser} across a pool of solutions that were generated throughout the competition time frame, considering the bonus coefficients at the time of submission. The solution, submitted on July 16, 2023 at 01:23 PM, UTC, consists of 27 ships that mine $17,614.26$ kg of material from 222 asteroids. At the time of submission, this resulted in a bonus-penalized score of $15,488.896$, which was enough to place 5th in the overall competition. 

In the submitted solution, the bonus coefficients did not have any impact on the optimal solution set. This is because the submitted solution lies very close to the cutoff permitting 27 ships; alternative combinations in an attempt to mine asteroids with better bonus coefficients simply are not included because the benefit of having another ship outweighs these losses. No gravity assists are used throughout all ship trajectories. Our analysis indicated that they tended to be too detrimental to the time of flight to provide an advantage. 

Our pruning cuts combined with the subset generation algorithm (described in Section~\ref{sec:Subsets}) resulted in almost all trajectories rendezvousing asteroids with semi-major axis between $2.7$ and $3.0$ AU. A clear outlier is the 7 asteroid trajectory returning $613.00$ kg of mined mass, which rendezvouses asteroids with semi-major axis of around $2.2$ AU. The subset used for this solution is derived from the graph theory approach; in particular, it is the only solution outside of the main block seen in Figure~\ref{fig:clusteringCri}. 
\begin{figure}[hbt!]
    \centering
    \includegraphics[width = 16cm]{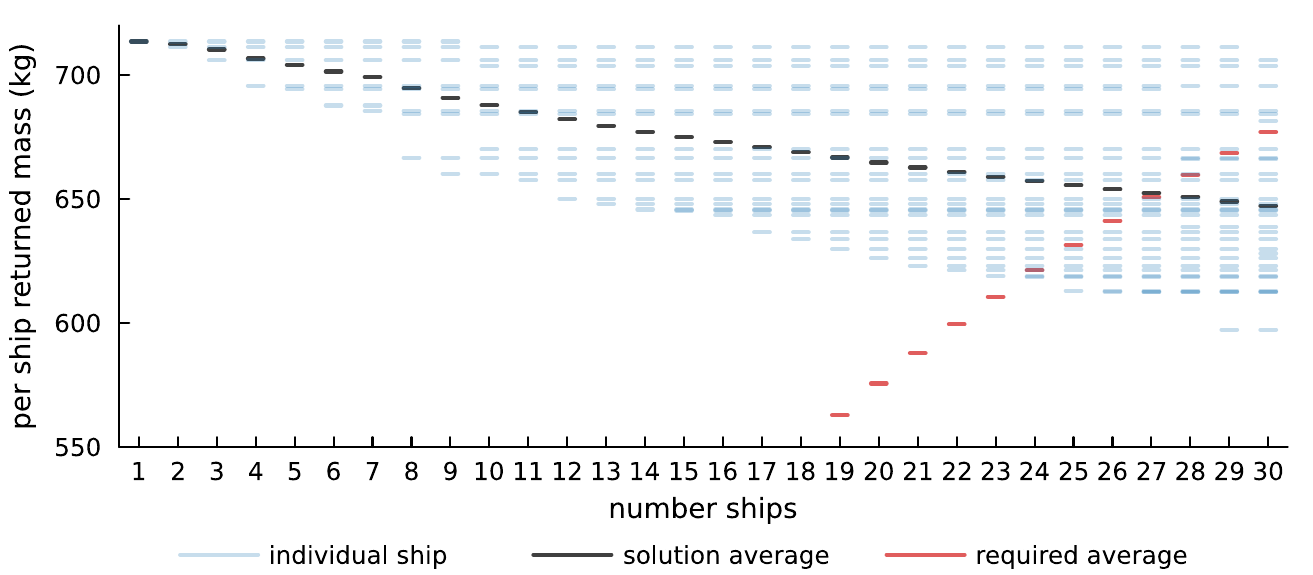}
    \caption{Results of the optimal selection algorithm parameterized by the maximum number of ships $N$. The average per ship returned mass for our submitted solution is only above that permitted for number ships $\leq27$.}
    \label{fig:best_combinations}
\end{figure}

An interesting visualization is provided in Figure~\ref{fig:best_combinations}. This demonstrates the combinations of self-cleaning ships that would be used to form solutions of various sizes. It is clear that a greedy selection approach would have been suboptimal as some of the best-scoring self-cleaning solutions are removed (notably between 9 and 10 ship combinations) to allow for the introduction of other ships that give a higher score in combination. The best single self-cleaning ship in the solution pool returned $713.57$ kg from 9 asteroids but was not included in the submitted solution due to this reason. 

In many cases, such as in most of the $\ge 700$ kg mined mass returned ships presented in our solution, rendezvousing with 9 asteroids seems to be the practical limit for creating optimal self-cleaning sequences. Post-competition analysis has demonstrated that this is not necessarily the case. While searching in an individual subset, not only was a $763.30$ kg mined mass 9 asteroid chain found but also a $780.81$ kg mined mass 10 asteroid chain. This would theoretically indicate the possibility of solutions with 45 ships, but due to the unlikelihood of many independent sequences of this quality, it would not be expected to be possible. In searching for similar high-scoring self-cleaning solutions after the competition and performing additional optimization processes on the self-cleaning solutions submitted by the top 10 teams, the authors have constructed a solution consisting of 37 self-cleaning ships that mine a total of $27,045$ kg of material ($730.95$ kg per ship). In a similar search, including mixed solutions, the authors constructed a solution consisting of 39 ships that mine a total of $28,975$ kg of material ($742.95$ kg per ship). Importantly, this solution did not consist of only mixed-ship strategies but contained 17 self-cleaning ships. This demonstrates that self-cleaning strategies, while significantly reducing the search space and thus simplifying the optimization problem, may not have as large a drawback on optimality as many teams (including us) might have thought. Indeed, we have demonstrated the possibility of winning the competition using a self-cleaning solution strategy.

\begin{figure}[hbt!]
    \centering
    \includegraphics[width = \textwidth]{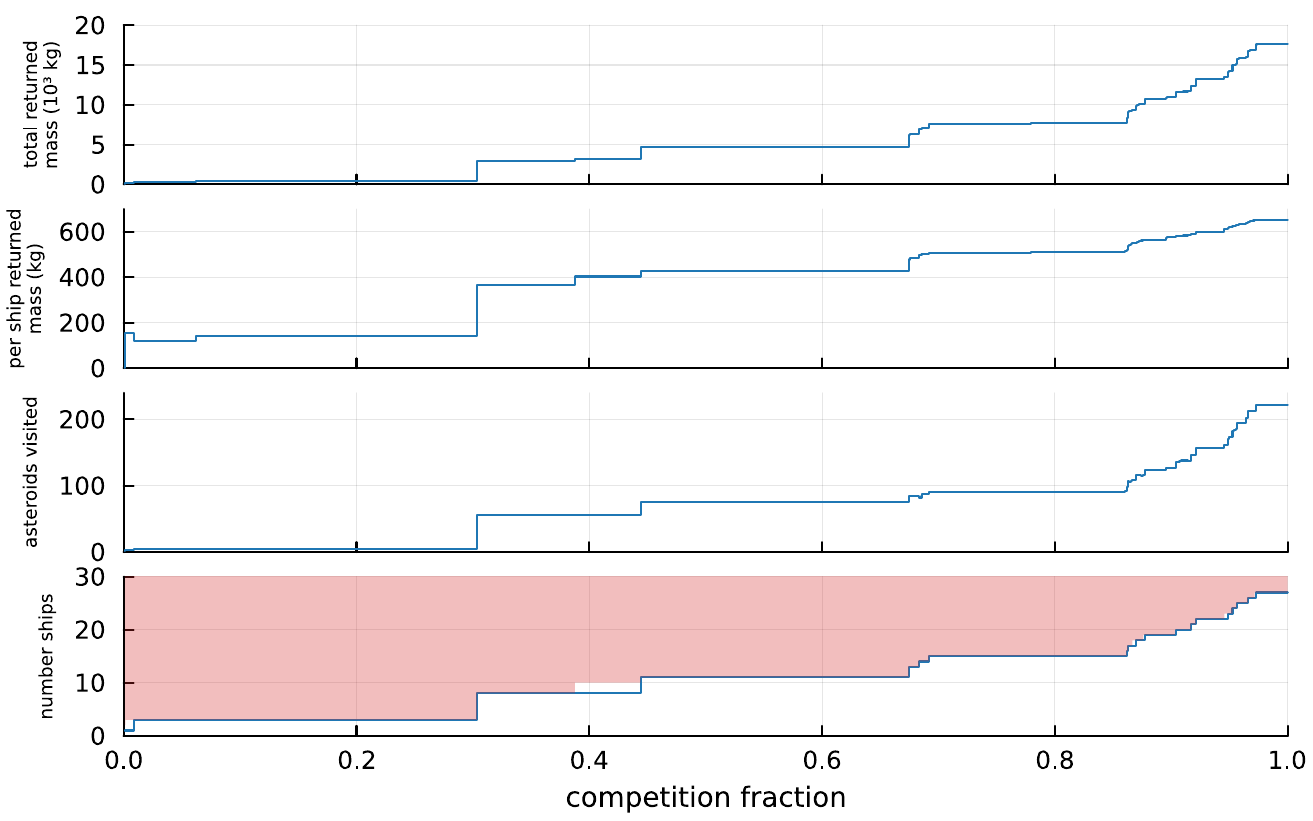}
    \caption{The evolution of a few key statistics of our best solution throughout the competition duration. For the bottom panel, the red region represents the area infeasible due to the ship number limit constraint.}
    \label{fig:ScoreAgainstTime}
\end{figure}

The timeframe of the GTOC 12 problem is perhaps the most difficult part of such competitions. Our solution dramatically increased in quality over the final few days as different parts of our final process came together. This process is visualized in Figure~\ref{fig:ScoreAgainstTime}, demonstrating several important statistics on the teams' performance throughout the competition. The number of ships permitted (and included) in the final submitted solution was almost double that of a week prior. A primary reason for this improvement was the introduction of the subset generation algorithm, which permitted greater numbers of high-quality, independent solutions from the beam search.

\section{\label{sec:Conclusion}Conclusions and Limitations}

This paper presents the work of the team 'TheAntipodes' throughout the timeframe of GTOC 12. Our initial expectations of challenging competition were confirmed, compelling our team to develop and innovate methodologies that may not have been seen in previous iterations of the GTOC competition. The manuscript reports key ideas and methodologies that contributed to our submitted solution. However, many attempts that did not succeed were not documented, while post-competition developments will be part of a separate publication. 

The decision to work on self-cleaning missions was instrumental in simplifying the problem and producing feasible solutions rapidly. The introduction of asteroid `subsets' permitted the high performance of a beam search tailored for the problem, and the introduction of a heuristic helped the impulsive to low-thrust conversion. Solving the low-thrust trajectory optimization problems did not present any computational challenges, thanks to the introduction of sequential convex programming. Moreover, the computational efficiency of this approach opens the door to the potential development of a beam search method that works with accurate low-thrust arcs. Manual refinement of the rendezvous times permitted several more ships to be included in the final submitted solution.

While we have demonstrated the effectiveness of self-cleaning missions in achieving very good solutions, this strategic choice represents a limitation in our approach. Secondly, although mitigated by the developed heuristic, the utilization of Lambert's arcs in the beam search for establishing asteroid sequences posed challenges in the convergence of the low-thrust problem. Lastly, the tight competition schedule precluded the development of a fully automated method for optimizing entire missions at once. Post-competition, the completion of this automated approach revealed the sub-optimality of manual refinement and showcased the immense potential of sequential convex programming in optimizing complex multi-target scenarios, which will be discussed in an upcoming publication.

\section*{Acknowledgments}

The authors would like to thank: (1) Dr. Joan-Pau Sánchez Cuartielles and Jose-Carlos García Mateas for their discussions and feedback, which provided food for thought throughout the competition, (2) Xin Liu and Yashdeep Chaudhary for their work in the initial analysis of the problem and attempts to find better approximations to low-thrust transfers than Lambert transfers and (3) Alberto Fossa for their efforts to build a framework to run Julia code from MATLAB.

\bibliographystyle{elsarticle-num} 
\bibliography{library.bib}   

\end{document}